\documentclass[letterpaper,twocolumn,10pt]{article}
\usepackage{usenix2019_v3,epsfig,endnotes}
\usepackage{times, algorithm, algpseudocode}
\usepackage{url}
\usepackage{graphicx,epstopdf}
\usepackage{subfigure}
\usepackage{enumitem}	
\usepackage{xcolor}
\usepackage{amsmath}
\usepackage{listings}
\usepackage{changepage}
\usepackage{pifont}
\usepackage{multirow}
\usepackage{wrapfig}
\usepackage{color,soul}
\usepackage{tikz}
\usepackage{booktabs}

\newcommand*\circlew[1]{\tikz[baseline=(char.base)]{
		\node[shape=circle,fill=white,draw,text=black,inner sep=1pt] (char) {#1};}}

\newcommand{\Paragraph}[1]{\vskip 3pt\noindent\textbf{#1 }}	

\newcommand{\cmark}{\ding{51}}%
  {\begin{itemize}
	[leftmargin=0cm,
		itemindent=.3cm,
		labelwidth=\itemindent,
		labelsep=0pt,
		parsep=3pt,
		topsep=2pt,
		itemsep=1pt,
		align=left]
  }%
  {\end{itemize}}

\newenvironment{myenumerate}%
  {\begin{enumerate}
	[leftmargin=0cm,itemindent=.5cm,labelwidth=\itemindent,
		labelsep=0pt,
		parsep=1pt,
		topsep=1pt,
		itemsep=3pt,
		align=left]
  }%
  {\end{enumerate}}

\newcommand{\camera}[1]{#1}

\newcommand*{\affaddr}[1]{#1} 
\newcommand*{\affmark}[1][*]{\textsuperscript{#1}}

\newcommand{\name}{{Philly}}

\begin{document}

\date{}

\title{\Large \bf Analysis of Large-Scale Multi-Tenant GPU Clusters\\ for DNN Training Workloads}


\author{
\rm Myeongjae Jeon\affmark[\dag]\affmark[\**], \rm Shivaram Venkataraman\affmark[\ddag]\affmark[\**], \rm Amar Phanishayee\affmark[\**], \\
\rm Junjie Qian\affmark[\**], \rm Wencong Xiao\affmark[\S]\affmark[\**], and \rm Fan Yang\affmark[\**]\\
\\
\affaddr{\affmark[\dag]UNIST~~~~~~~\affmark[\ddag]University of
Wisconsin~~~~~~~\affmark[\S]Beihang University ~~~~~~~\affmark[\**]Microsoft Research}
}

\maketitle

\begin{abstract}
With widespread advances in machine learning, a number of large enterprises
are beginning to incorporate machine learning models across a number of
products. These models are typically trained on shared, multi-tenant GPU
clusters. Similar to existing cluster computing workloads, scheduling
frameworks aim to provide features like high efficiency, resource
isolation, fair sharing across users, etc. However Deep Neural Network
(DNN) based workloads, predominantly trained on GPUs, differ in two
significant ways from traditional big data analytics workloads. First, from
a cluster utilization perspective, GPUs represent a monolithic resource
that cannot be shared at a fine granularity across users. Second, from a
workload perspective, deep learning frameworks require gang scheduling
reducing the flexibility of scheduling and making the jobs themselves
inelastic to failures at runtime. In this paper we present a detailed
workload characterization of a two-month long trace from a multi-tenant GPU
cluster in \camera{Microsoft}. By correlating scheduler logs with logs from
individual jobs, we study three distinct issues that affect cluster
utilization for DNN training workloads on multi-tenant clusters: (1) the
effect of gang scheduling and locality constraints on queuing, (2) the
effect of locality on GPU utilization, and (3) failures during training.
Based on our experience running a large-scale operation, we provide design
guidelines pertaining to next-generation cluster schedulers for DNN
training workloads.
\end{abstract}


\section{Introduction}
\label{sec:Introduction}

Recent advances in machine learning have led to tremendous improvements in
tasks ranging from object detection~\cite{alexnet} to speech
recognition~\cite{may2017kernel} and language
translation~\cite{vaswani2017attention}. As a result a number of enterprises
are now incorporating machine learning models in various
products~\cite{google-photo-deep,siri-deep}. To facilitate model training,
enterprises typically setup a large cluster shared by users belonging to a
number of different production groups. Similar to clusters setup for big data
analysis~\cite{Verma15, chaiken2008scope}, using shared clusters can
facilitate better utilization and reduce development overheads.

However deep learning workloads pose a number of new requirements or
constraints on cluster management systems. Since machine learning algorithms
are floating point computation intensive, these workloads require hardware
accelerators like GPUs. However, unlike CPUs, accelerators do not typically
have proper hardware support for fine-grained sharing~\cite{Juncheng19}.
While there are software mechanisms to enable sharing, they often have high
overhead making it challenging to share resources across
jobs~\cite{Yu17,Rhu16}. Furthermore, training on large datasets often
requires the use of multiple GPUs~\cite{goyal2017accurate} and machine
learning frameworks typically require that tasks on each GPU be scheduled at
the same time, i.e., gang scheduled~\cite{feitelson1996packing}. This
increases the risk of resource fragmentation and low utilization in shared
clusters. Finally, multi-GPU training also implies synchronization of model
parameters across GPUs and hence it is important to achieve better
\emph{locality} while scheduling to allow for the use of faster interconnects
for both intra- and inter-machine communication.

Despite their growing popularity, to the best of our knowledge, there has
been no systematic study of multi-tenant clusters used to train machine
learning models. In this paper, we present the design of a large,
multi-tenant GPU-based cluster used for training deep learning models in
production.
We describe {\name}, a service \camera{in Microsoft} for training machine
learning models that performs resource scheduling and cluster management for
jobs running on the cluster. Using data from this system, we then present a
detailed workload characterization and study how factors such as gang
scheduling, locality requirements and failures affect \emph{cluster
utilization}.


Our analysis spans across two months and uses around 100,000 jobs run by
hundreds of users. We combine logs from Apache YARN~\cite{yarn}, our cluster
scheduler, utilization information from Ganglia~\cite{ganglia}, and logs from
each job to perform a systematic analysis of cluster utilization.

We study two main aspects of how locality-aware scheduling affects performance and utilization.
First, we study how waiting for locality constraints can influence queuing delays before training jobs are run.
Training jobs need to be
gang scheduled, as hyper-parameters are picked for specific GPU count
configurations. Given that training jobs take a long time to run, and greater
locality improves performance due to the availability of faster interconnects
for parallel training~\cite{wencong1}, the scheduler in {\name} waits for
appropriate availability of GPUs before beginning to run the training job. Our study
shows that as one might
expect, relaxing locality constraints reduces queueing delays, especially for
jobs that use many GPUs \--- our emphasis here is not on presenting this as a
new insight, but instead on highlighting this using real-world data from
production clusters.

Next, we study how locality-aware scheduling can affect the GPU utilization
for distributed training jobs. 
Even though most GPUs within a cluster are allocated to users, thus
\textit{suggesting} high cluster utilization, this metric alone is
misleading. We show that the hardware utilization of GPUs in use is only
around 52\% on average. We investigate two reasons which contribute to low
GPU utilization: (1) the \emph{distribution} of \textit{individual} jobs
across servers, ignoring locality constraints, increases synchronization
overheads, and (2) the \emph{colocation} or packing of \textit{different}
jobs on same server leads to interference due to contention for shared
resources.

Finally, we look at why jobs might fail to complete successfully and offer a
detailed characterization of the causes for such failures in our clusters.
Around 30\% of jobs are killed or
finish unsuccessfully due to failures. Failures are caused by errors across
the stack, with programming errors dominating failures and occurring early in
the training process; failures due to cluster components like HDFS tend to
occur much later in the training lifecycle.


Based on the lessons learnt from data analysis and our experiences running a
large-scale operation over the years, we provide three guidelines to improve
the next generation of cluster schedulers for DNN workloads. First, because
the lack of locality impacts both utilization and job runtime, and because
DNN training jobs are long running, schedulers should trade queueing delay
for adhering to locality constraints. Second, different jobs that share a
single server may interfere with each other and thus adversely affect their
training time. Schedulers should thus aim to isolate the jobs on dedicated
servers while implementing techniques like migration for defragmentation,
to support the locality constraints of jobs that need more GPUs. Third, many
failures ought to be caught early, well before they are scheduled on a larger
shared cluster. This can be achieved by scheduling each incoming job on a
small dedicated pool of servers or even using a single GPU should be able to
catch simple programming and configuration errors from multi-GPU jobs.
Furthermore, an online analysis of failures at runtime can let schedulers
adapt their retry policies thus avoiding wasteful re-execution.


\camera{Philly’s design does not stand in isolation. There are many open
platforms for DNN job scheduling that use designs similar to {\name}, e.g.,
OpenPAI~\cite{openpai} and Submarine~\cite{submarine}.} We hope that insights
and data from our study, and the accompanying traces~\cite{phillytrace},
inform the burgeoning work of scheduling research for machine learning
workloads.

\section{{\name}: System Overview}
In this section we provide an overview of the design and architecture of
{\name}. First, we describe the workloads that are supported in our system
and then describe the hardware characteristics of the clusters. Next, we
describe the lifecycle of a job. Finally, we explain our data collection
pipeline and highlight the data we use to perform our analysis in subsequent
sections. The authors would like to note that {\name} has been developed over
the past few years by a team of developers in our company and has gone
through multiple generations of design.


\subsection{Workloads}
Philly is designed to support workloads that perform supervised machine
learning where jobs learn a model given training data and the corresponding
labels. This includes training jobs from production groups developing
products that use models for image classification, speech recognition, etc.
The system supports jobs written using any machine
learning framework like TensorFlow~\cite{tensorflow}, CNTK~\cite{cntk},
Caffe~\cite{caffe}, and PyTorch~\cite{PyTorch}. Jobs are based on recently
proposed learning architectures like convolutional neural
networks~\cite{alexnet}, LSTMs~\cite{lstms} and RNNs~\cite{rnns}.

All jobs, irrespective of the framework or model being used, rely on
iterative optimization methods~\cite{goodfellow2016deep} like stochastic
gradient descent (SGD). In each iteration,
the gradient computation is performed by translating the model components
into
code that can be
executed on accelerators like GPUs. The gradient values are then aggregated
to compute a model update and these iterations are repeated until
convergence. Training a model could require thousands to millions of
iterations~\cite{inception}, and result in multiple passes or \emph{epochs}
over the entire dataset.

To scale training across larger datasets, a number of jobs use distributed
training across machines. Distributed training typically uses data
parallelism where each worker loads a complete copy of the model into its own
memory. In each iteration, every worker performs training using a subset of
the input data, and at the end of the iteration all workers exchange
gradients to synchronize model updates.  This synchronization phase is
performed using either \camera{parameter servers~\cite{osdi14ps} or high
performance libraries for collective communication (such as MPI, NCCL, etc).}

\subsection{Cluster Architecture}
Our system is deployed on large GPU clusters shared across many groups in the
company. Our clusters has grown significantly over time, both in terms of the
number of machines ($5\times$ increase in one year) as well as the number of
GPUs per machine (2-GPU to 8-GPU servers).

Our clusters have high-speed network connectivity among servers and GPUs in
the cluster. This is to speed up distributed training where workers need to
exchange model updates promptly for every iteration. There is a hierarchy of
network links available in our cluster for communication across GPUs. For
example, machines within the same rack (\textit{RDMA domain}) are connected
via 100-Gbps RDMA (InfiniBand) network, while cross-rack traffic goes through
Ethernet. To improve communication performance, workers in a distributed
training job must either be colocated on the same machine or preferably
communicate over a higher-speed network such as say InfiniBand. Thus, our
framework considers both GPUs and network connectivity for scheduling.

Similar to existing big data analytics clusters, our clusters use
HDFS~\cite{hdfs} as the distributed storage system and our resource manager
is based off Apache YARN~\cite{yarn}. Input data for the machine learning
jobs is stored in HDFS and read by jobs during training.
\camera{Users provide a Docker container with their training code and its
dependencies. Each training job requests 1 or more GPUs which can be
allocated across multiple machines. {\name} instantiates one container per
machine allocated to the job when it is scheduled for execution.}

\begin{figure}[!t]
\center
\includegraphics[width=3.3in]{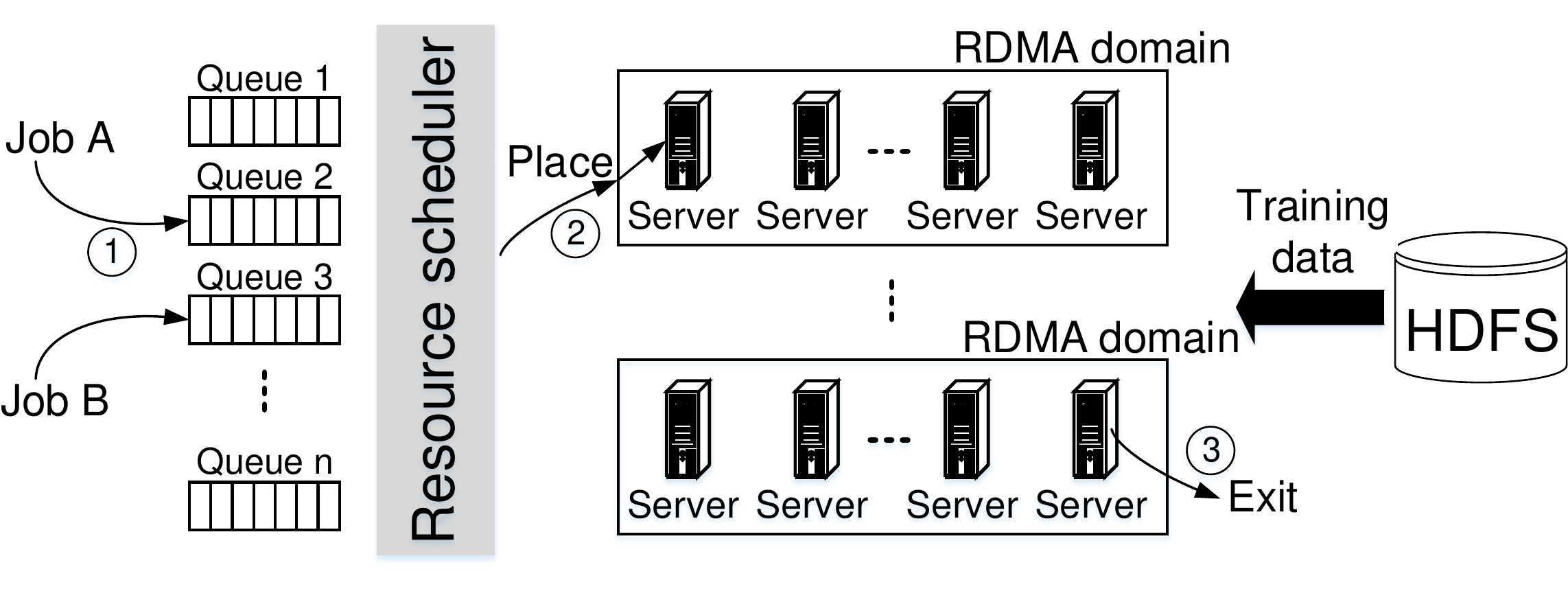}
\caption{\label{fig:arch} The lifecycle of deep learning jobs in {\name}.}
\end{figure}

\subsection{Job Scheduling and Execution Workflow}
Figure~\ref{fig:arch} shows the lifecycle of a deep learning job in {\name}
and the various stages of execution that it goes through.

\Paragraph{Incoming jobs and queueing \circlew{1}.}
As a part of job submission, users specify \emph{the number of GPUs}
required. To facilitate host resource allocation, \camera{we perform an
allocation of CPU cores and memory capacity proportional to the requested GPU
count.}
Once a job has been received by the scheduler it is
queued while the necessary GPUs are allocated. To support multiple production
groups we create a virtual cluster for each group and associate a
\emph{resource share} or \emph{quota} in terms of number of GPUs to each
virtual cluster. Each virtual cluster has a separate allocation queue in
Apache YARN and we use the Fair Scheduler to manage these
queues~\cite{yarnfairscheduler}. Our scheduler not only respects the
configured resource shares but also allocates unused GPUs to a queue which
has additional demand. Jobs can be preempted based on fair share of resources
among virtual clusters. Our scheduler starts preemption only when a majority
(90\%) of total GPUs are being used.

For distributed learning, deep learning frameworks require all the GPUs to be
available at the same time~\cite{Juncheng19}. Thus the scheduler needs to
perform \emph{gang scheduling} while being \emph{locality-aware}, i.e., pack
a job's GPUs onto the smallest number of servers and within an RDMA domain.
Locality awareness improves training time by bringing down the time needed
for parameter synchronization~\cite{wencong1,Juncheng19} due to the
availability of: (i) fast intra-server interconnects (such as PCIe and
NVLink), and (ii) for jobs that do not fit on a single server, high-bandwidth
links available within an RDMA domain.
We implement these goals by acquiring resources for a job as GPUs become
available and waiting for a pre-specified timeout (2--3 minutes in our setup)
to acquire all the necessary GPUs with the locality constraints. To
facilitate locality-aware GPU scheduling, our job scheduler keeps track of
all idle GPUs in the cluster and ranks the corresponding racks and servers.
Specifically, racks are ranked by increasing order of allocation or
occupancy, and the machines in a rack are ordered the same way. This allows
the scheduler to first consider racks and then servers within those racks
that have most GPUs available.

If the request is not fulfilled by the timeout, any partially acquired
resources are relinquished and we retry scheduling after a back-off (2
minutes in our setup). To avoid starvation, the locality constraints are
relaxed after a scheduling request has been retried a fixed number of times.
We analyze corresponding queuing delays in
Section~\ref{sec:job_queueing_analysis}.


\Paragraph{Job placement and utilization \circlew{2}.} While the scheduler
tries to maximize locality for distributed jobs as described before, at the
same time the scheduler also aims to avoid fragmentation of resources from
smaller jobs (e.g., 1-GPU jobs) by packing them into a fewer servers. However
colocating different jobs on the same server could lead to lower GPU
utilization due to interference in shared system resources such as PCIe
bus~\cite{wencong1}. In order
to better understand this trade-off we study the effects of colocation vs.
distribution and measure how that affects utilization.

\begin{table*}[!t]
  \centering
  \caption{Comparison of DNN cluster schedulers. JCT means job completion time.}
  \label{tab:solution-summary}
  \small
  \begin{tabular}{l||c|c|c|c}
    \hline
     & \name{} & Gandiva~\cite{wencong1} & Optimus~\cite{Peng18} & Tiresias~\cite{Juncheng19} \\
    \hline
    Objective & Consolidation & Consolidation & Average JCT & Average JCT\\
    Algorithm & Locality-based  & Time-sharing & SRTF & Gittins Index \& LAS\\
    Input & Arrival time & N/A & Remaining time & Attained service\\
    Preemption & Model checkpoint & Context switch & Model checkpoint & Model checkpoint\\
    \hline
  \end{tabular}
\end{table*}

Once the job is scheduled to run, its GPUs are \emph{not shared} with other
jobs. This is because model training can be computation intensive and we need
consistent performance among workers of the job without having stragglers.
However, dedicated GPUs may be underutilized for many reasons, e.g.,
inefficiencies in the code generated by the machine learning frameworks or
programs blocking on I/O when reading data from storage. GPU underutilization
also comes from distributed training where computation may block during model
synchronization among the workers. We analyze the effects of job placement
and GPU utilization in Section~\ref{sec:GPU_utilization_analysis}.

Table~\ref{tab:solution-summary} qualitatively compares \name{} with the
state-of-the-art DNN cluster schedulers, showing both similarities and
differences exist. Nonetheless, locality and colocation are the common issue
for all contemporary clusters, and that insights obtained in this study are
widely valuable.

\Paragraph{Training progress and completion \circlew{3}.} Jobs can finish
with one of three statuses: passed, killed, or unsuccessful. Passed indicates
that the job completed successfully, while killed indicates that the job was
terminated by the user.

Among successful jobs, every job runs a number of iterations to improve the
model incrementally, and the number of iterations to run is typically a
static parameter set by the user. In cases where a job is configured with too
many iterations, it is possible to deliver the same (or similar) quality of
trained model with fewer iterations.
Failed jobs in our system are retried a fixed number of times. This is useful
for overcoming non-deterministic failures and if the job does not succeed
after retries then it is marked as unsuccessful. As failures also contribute
to ineffective cluster utilization, we perform a detailed study to understand
the reasons behind failures in Section~\ref{sec:JobFailures}.

\camera{While our focus in this section is specifically about the lifecycle
and execution flow in {\name}, there are many open platforms for ML job
scheduling that use a similar design. Platforms like OpenPAI~\cite{openpai}
and Submarine~\cite{submarine} also use a centralized scheduler with support
for running machine learning frameworks as Docker containers. While the
details of the scheduling algorithm vary across systems, a number of aspects
we study in this paper are independent of the choice of scheduler: e.g.,
failures due to programming errors and bugs in popular frameworks, effect of
distributed training across machines, etc. Thus, we believe that lessons from
Philly are generally applicable to other clusters as well.}

\subsection{Data Collection and Analysis}
\label{sec:DataCollection} \camera{The cluster under study consists of
hundreds of machines accounting for thousands of GPUs of the same model. The
cluster has 2 server SKUs -- one with 2 GPUs per server and another with 8
GPUs per server; RDMA domains are homogeneous with respect to server SKUs.}
To get a comprehensive understanding of the characteristics of our system and
workloads, we developed a data collection and analysis pipeline and collect
logs over a 75-day period from Oct. 2017 to Dec. 2017. Our logs contain a
total of 96260 jobs over 14 virtual clusters.

The analysis pipeline combines three main log sources in our system as
follows. (1) We collect the YARN scheduler logs to obtain job arrival time,
number of GPUs requested, GPU allocation status, and job finish status. (2)
We collect stdout and stderr logs from the machine learning frameworks that
execute scheduled jobs.
(3) We collect logs from Ganglia monitoring system that reports per-minute
statistics on hardware usage on every server, including CPU, memory, network,
and GPU utilizations. Combined with GPU allocation status in YARN scheduler
logs, we can track how a scheduled job utilizes cluster hardware resources.


\begin{figure}[!t]
\center
\includegraphics[width=2.3in]{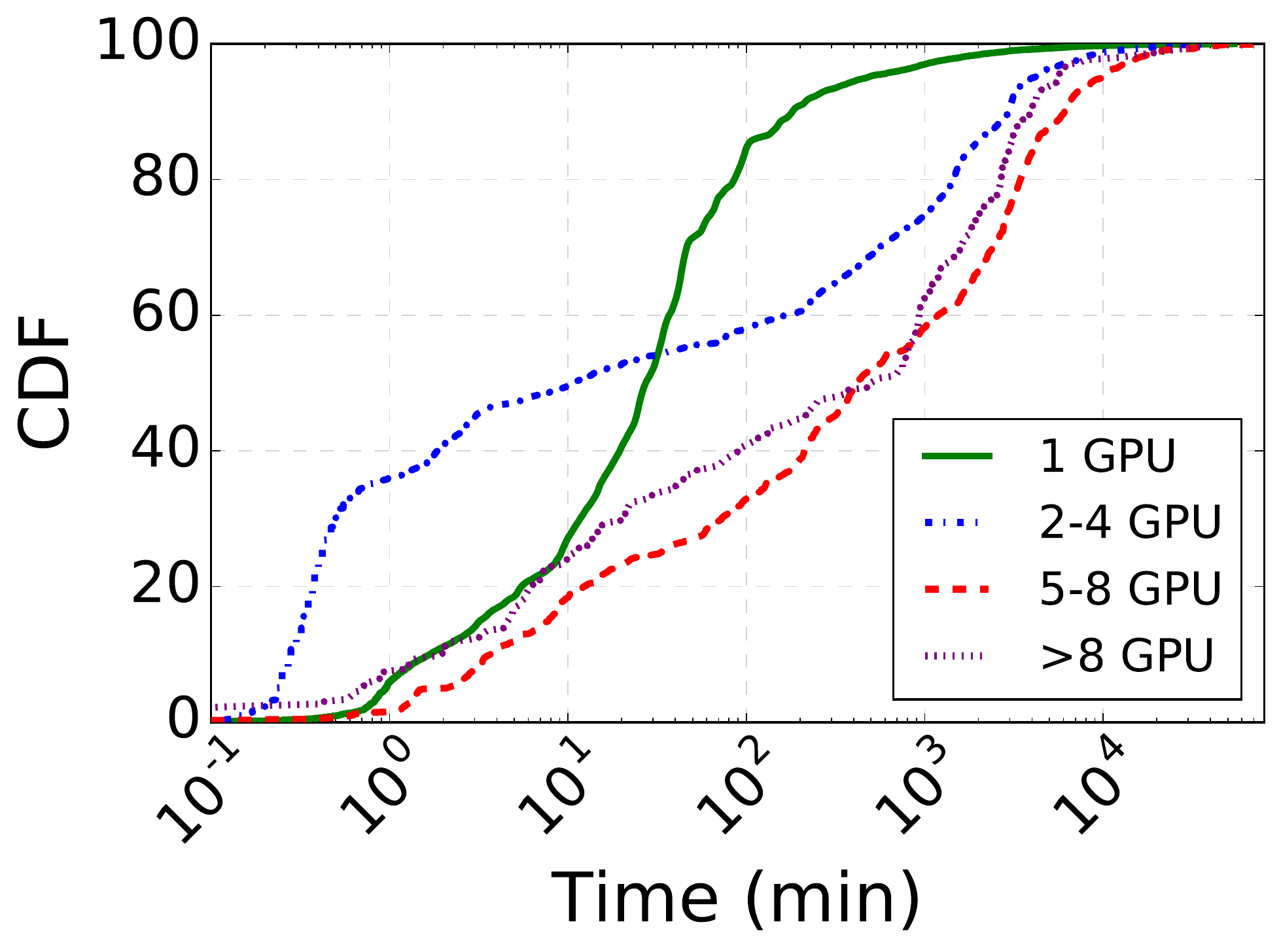}
\caption{\label{fig:job_basic} CDF of job run times for 1 GPU, 2-4 GPU, 5-8 GPU, and $>$8 GPU jobs.}
\end{figure}

Our collected data contains jobs from a wide spectrum in terms of their run
times and sizes, and consequently cluster resources demands. Jobs run from
minutes to days or even weeks, as shown in Figure~\ref{fig:job_basic}. In
contrast, in big data analytics, job execution times range from only tens of
milliseconds to a few hours~\cite{Ousterhout13,Schwarzkopf13,Boutin14}.
Furthermore, we see that our workload has significant skewness in run time,
with around 0.5\% jobs taking more than a week to be finished.
Figure~\ref{fig:job_basic} also shows how jobs of different sizes
vary in terms of execution times. We see that jobs with more GPUs tend to
run longer. This results in most of the cluster resources demands
coming from the larger jobs, and resource availability status changing
relatively slowly over time. 

\section{Impact of Locality Awareness}
\label{sec:Scheduler_impact}

\begin{figure*}[!t]
\centering
\subfigure[VC1]{\includegraphics[width=1.35in]{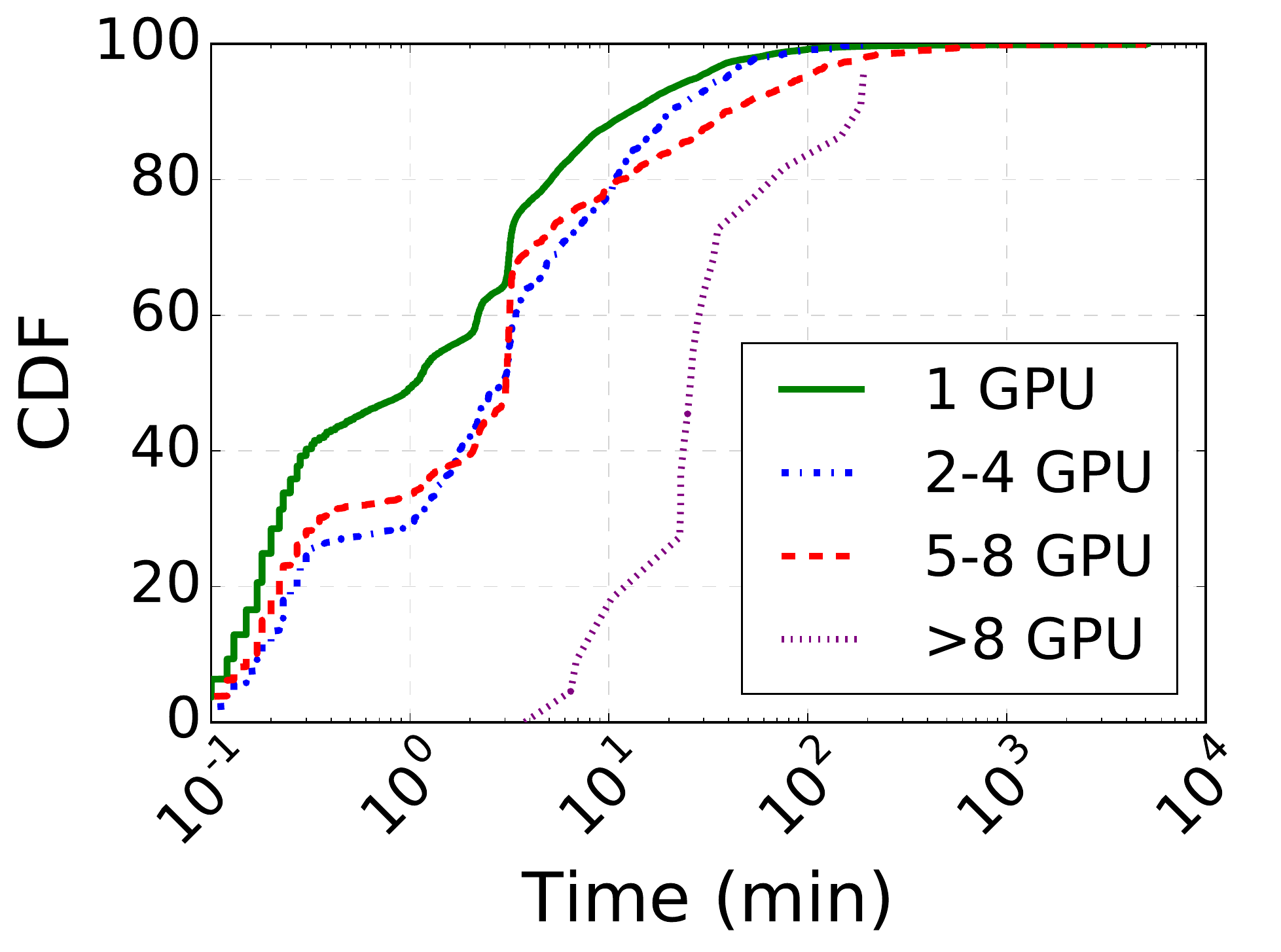}}
\subfigure[VC2]{\includegraphics[width=1.35in]{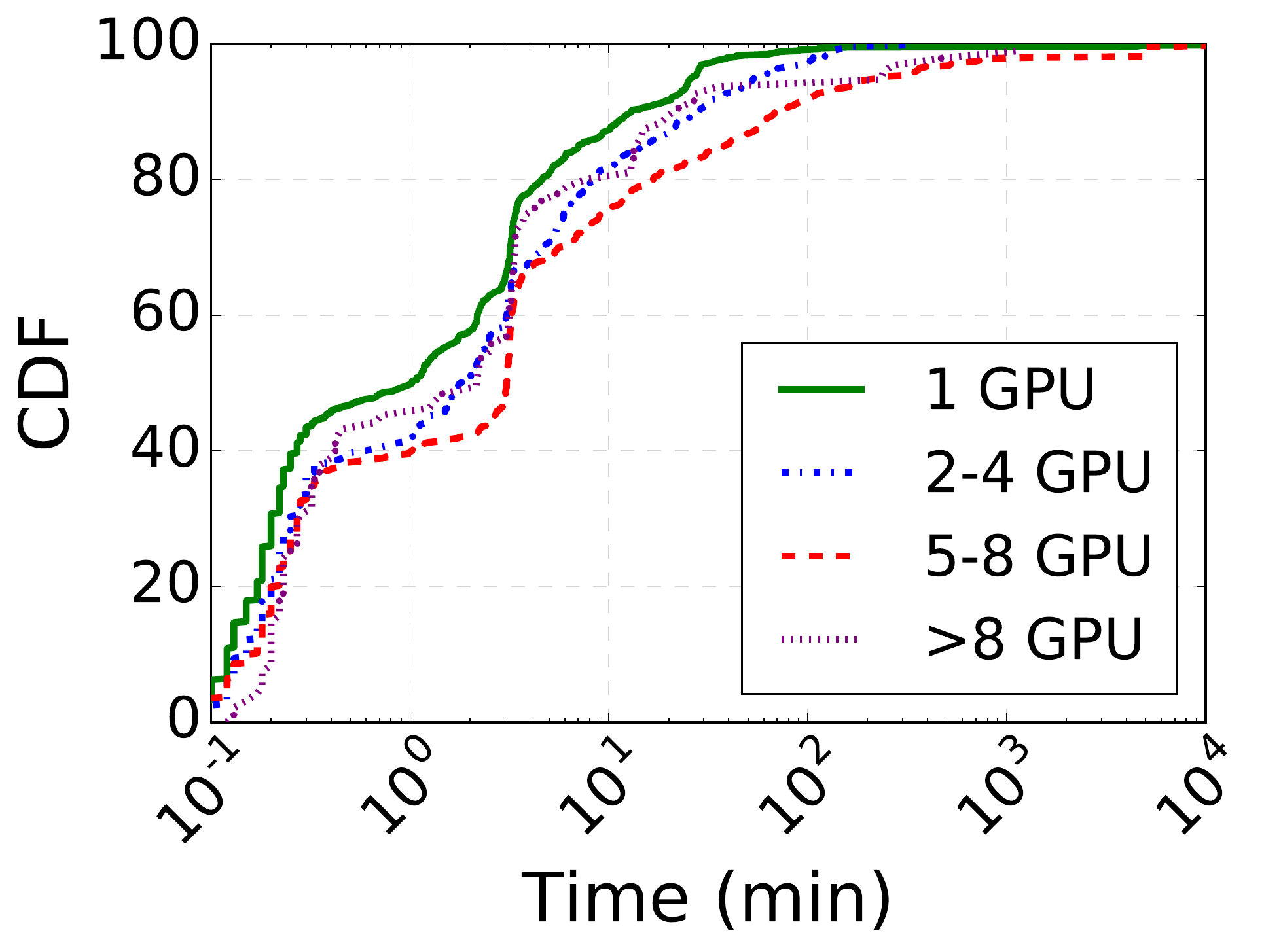}}
\subfigure[VC3]{\includegraphics[width=1.35in]{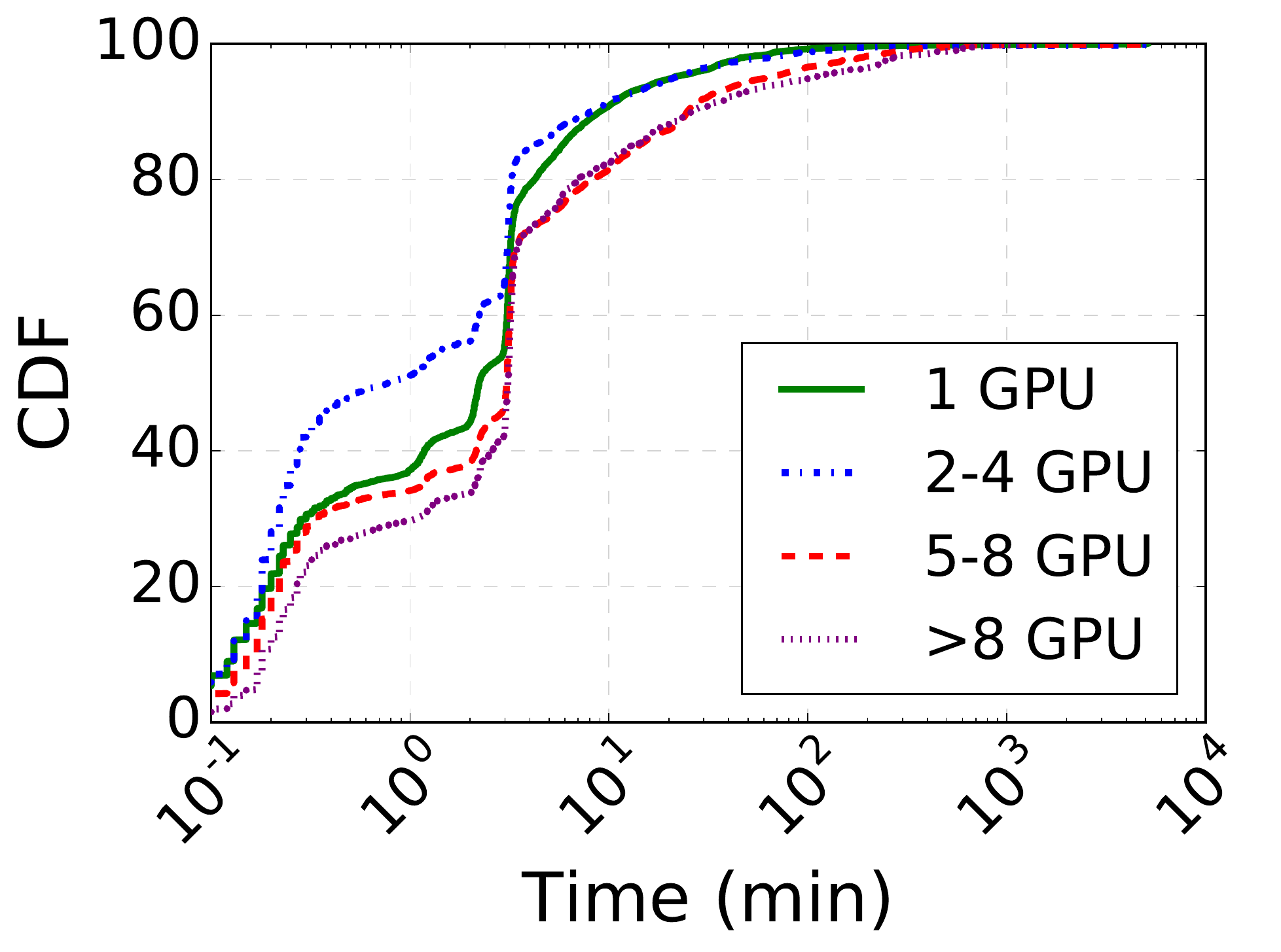}}
\subfigure[VC4]{\includegraphics[width=1.35in]{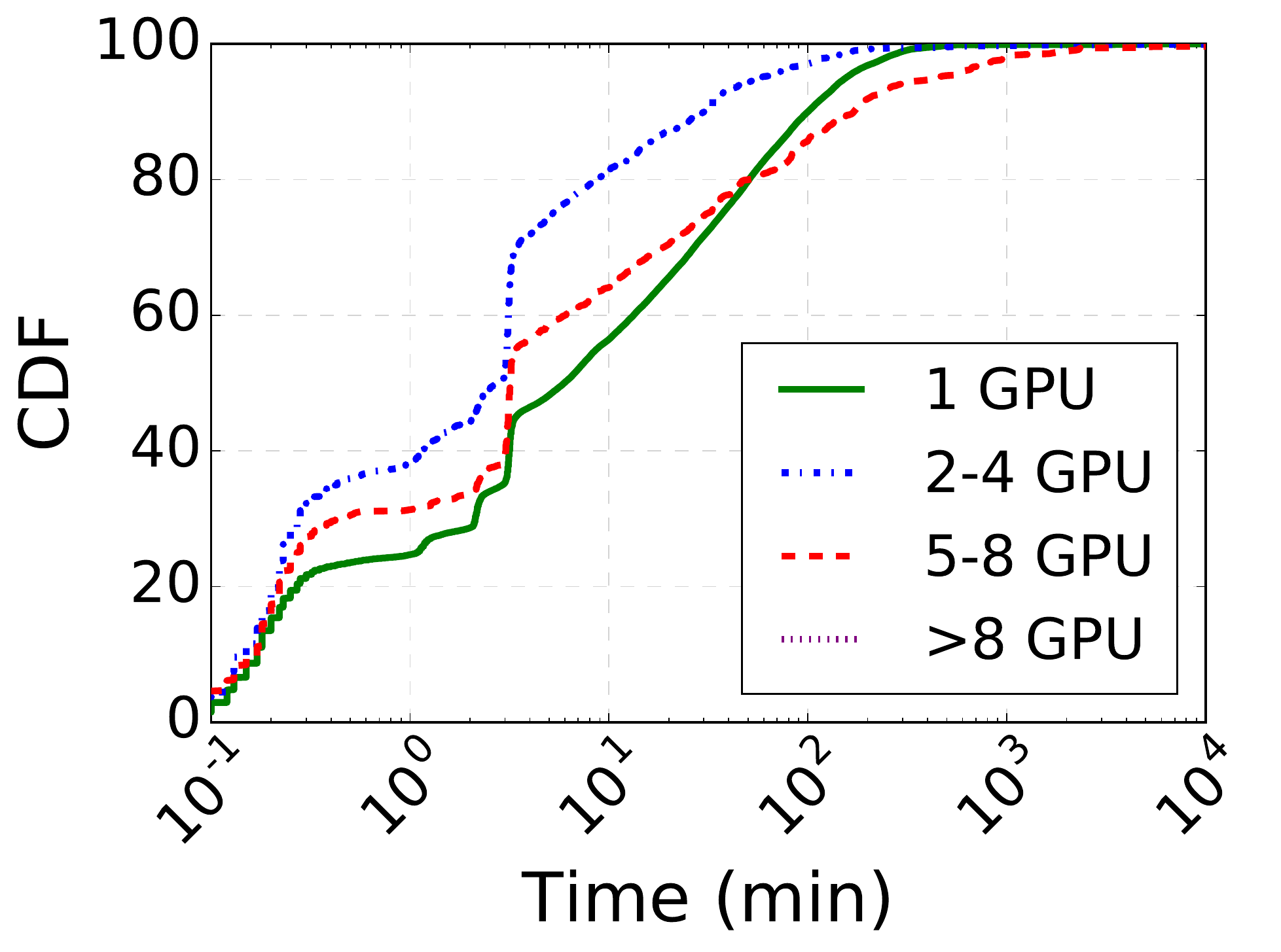}}
\subfigure[VC5]{\includegraphics[width=1.35in]{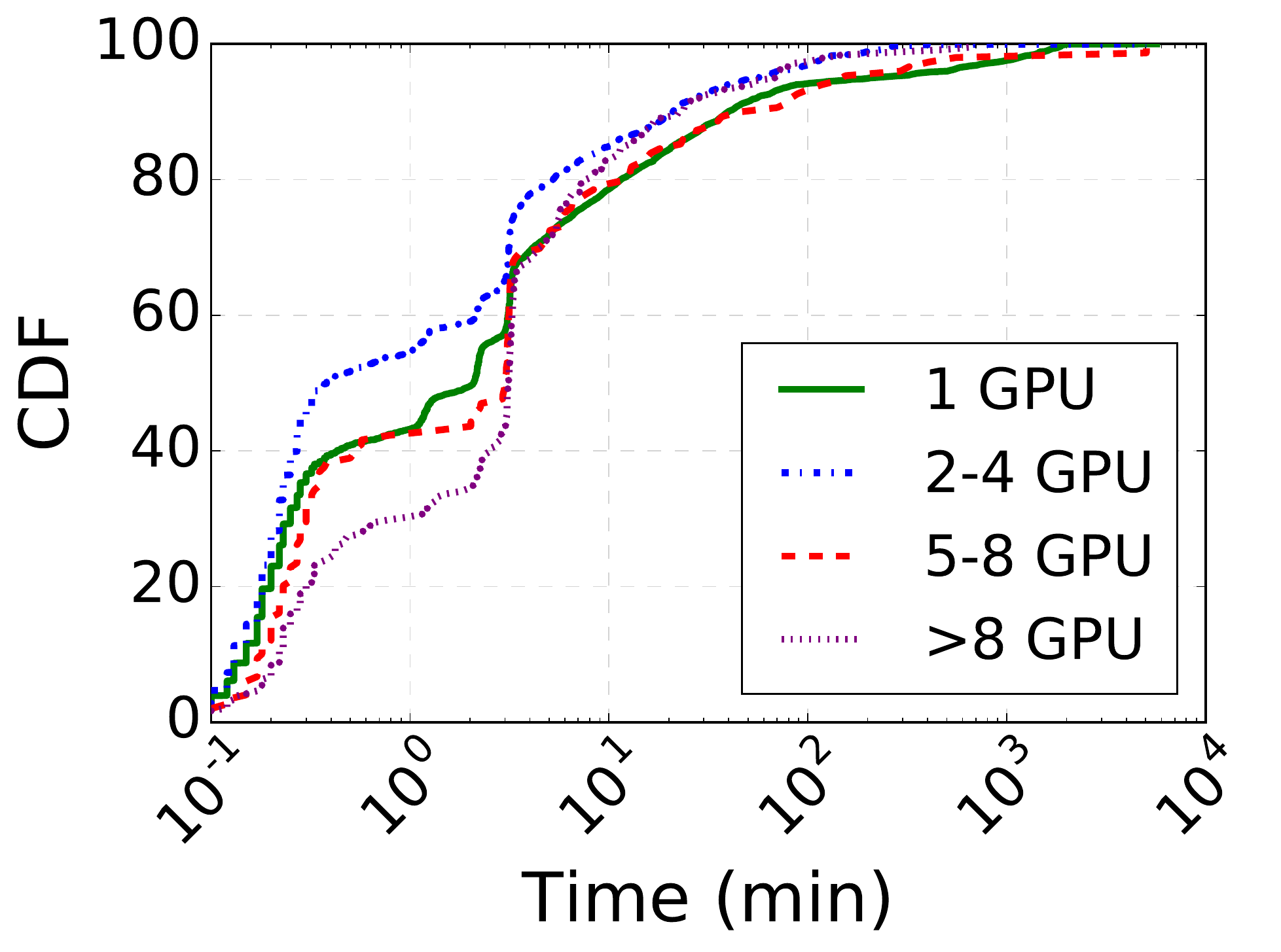}}
\caption{\label{fig:gpu_queuetime} CDF of scheduler queueing delay for five of the largest virtual clusters in our deployment.
\camera{Note that VC4 contains no jobs with $>$8 GPU.}}
\end{figure*}

Our scheduler trades off locality for lower waiting. Thus placement choices
made by the scheduler affect the efficiency of DNN training in two parts:
queueing delay (before job execution) and hardware utilization of in-use GPUs
(after job execution). The effect of locality constraints on queuing delays
has been extensively explored in large-scale resource
allocation~\cite{Isard09,Zaharia10,Ananthanarayanan11,Boutin14}. Machine
learning workloads introduce similar constraints driven by gang scheduling
and the requirement for using fast interconnects. In
Section~\ref{sec:job_queueing_analysis}, we analyze queueing delays in the
context of DNN training cluster using real-world data in detail. Next, we
study utilization of processing cycles for GPUs allocated to training jobs in
Section~\ref{sec:GPU_utilization_analysis}. In particular, while prior work
discusses efficiency of distributed training for a certain job size or a
configured placement~\cite{wencong1,Juncheng19}, we perform an analysis on
the aggregated efficiency for a range of job sizes for the first time.

\subsection{Queueing Delays} \label{sec:job_queueing_analysis}




We first consider overall queueing delay observed during job scheduling. We
plot the CDF of queueing delay in Figure~\ref{fig:gpu_queuetime} for all jobs
in five of the largest virtual clusters (VCs). Jobs that need more than 4
GPUs tend to have a slightly longer tail in the distribution of queueing
delays compared to their 1 GPU and 2-4 GPU counterparts.
For example for VC2, 25\% of jobs using $>$4 GPUs, \camera{which include both
5-8 GPU and $>$8 GPU,} experience a queueing delay of at least 10 minutes; in
comparison, only 10\% of 1 GPU jobs experience a queueing delay of at least
10 minutes.



\begin{figure}[!t]
\center
\includegraphics[width=2.1in]{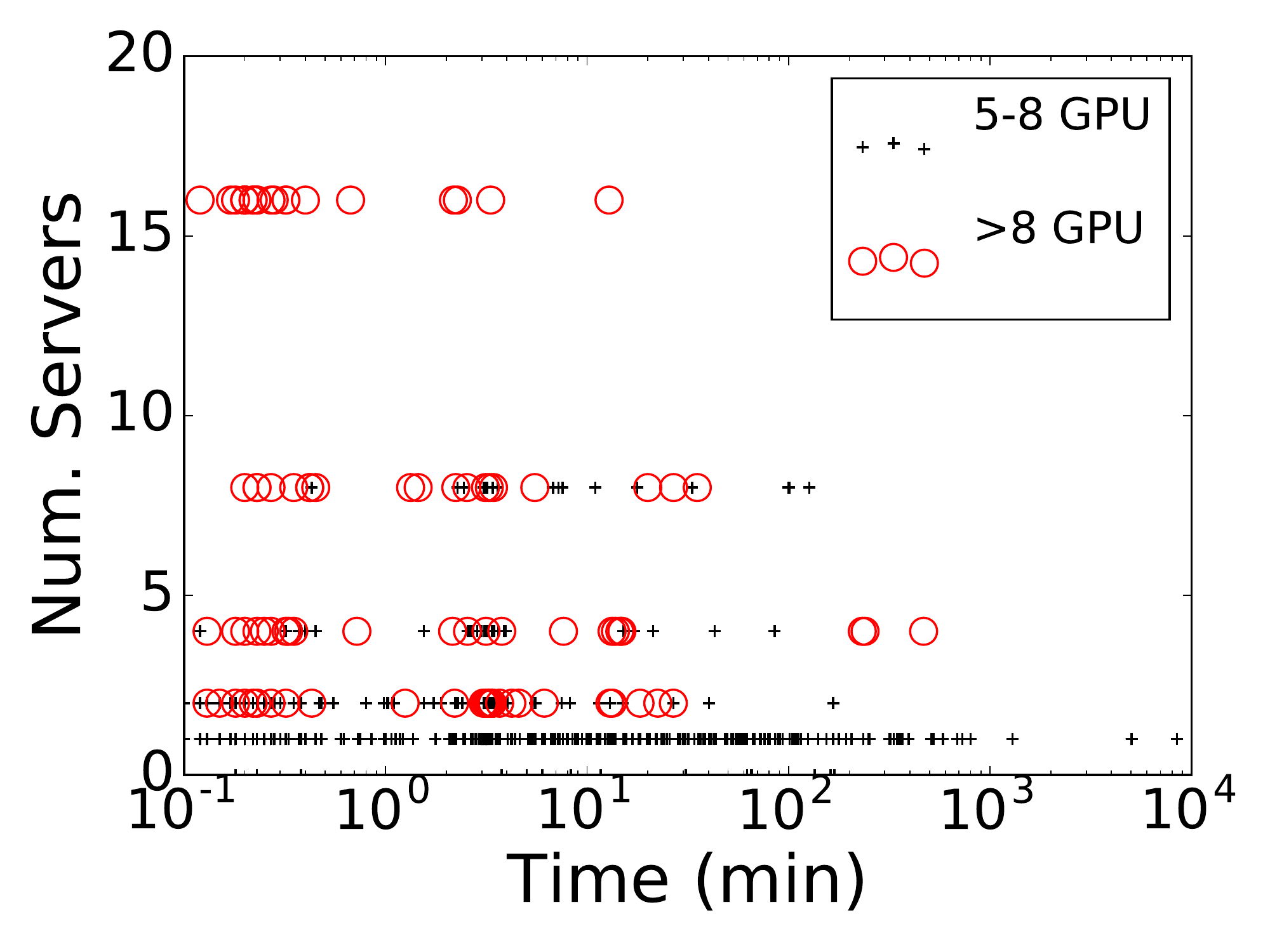}
  \caption{\label{fig:relaxed_locality_ipgimg} For a given GPU count, relaxing locality constraints reduces queueing delays (VC2).}
\end{figure}

But overall, queuing delays for jobs, irrespective of their GPU demand, are
not markedly distinct.
This is partially a consequence of our scheduling policy that chooses to
relax locality constraints in order to start a job without incurring a very
long queueing delay penalty. To highlight the relation between locality
constraints and queueing delays, we next consider \camera{jobs with 5-8 GPU
and $>$8 GPU.}
We correlate scheduler waiting times with number of servers on which the jobs
are placed, and show the results in Figure~\ref{fig:relaxed_locality_ipgimg}.
As expected, most of jobs with 5-8 GPU are scheduled with high locality,
i.e., placed on one or two servers. On the other hand, we find that jobs with
$>$8 GPU are spread across a wider range from 2 to 16 servers. Clearly, when
jobs end up running on 16 servers, they start execution much sooner than
running on 2 or 4 servers. This confirms how our scheduler works in practice
to trade-off locality for lower scheduling delay.

While effective, we find that this decision affects the GPU utilization as
discussed in Section~\ref{sec:GPU_utilization_analysis}.
We next look at more details on the queuing delay characteristics and break
down the delay by different causes.

\subsubsection{Impact of Locality-Driven Scheduling}

\label{sec:scheduler_impact} Queuing delay can be caused by two primary
factors: fairness (which is common in conventional data analytics clusters),
and locality requirement and resource fragmentation (which is more prevalent
in deep learning clusters). We call queueing caused by the first factor as
\emph{fair-share delay}, as it happens when the virtual cluster uses up its
assigned quota (i.e., number of GPUs). However, it is possible that a job
arrives within the quota but fails to be scheduled, mainly because resource
fragmentation makes it hard to find enough GPUs with high locality. We call
this queuing delay as \emph{fragmentation delay}. In practice, we find that
resource fragmentation is very common. For example, we observe that (i) when
two thirds of the total GPUs are being used, the fraction of servers that are
completely empty is less than 4.5\% and that (ii) these servers are spread
across RDMA domains.



\begin{table}[!t]
  \begin{adjustwidth}{-1in}{-1in}
    \centering
    \small
    {
      \begin{tabular}{c||c|c|c}
        \hline
        Delay & 2-4 GPU & 5-8 GPU & $>$8 GPU \\\hline
        Fair-share    & 5168 (40.6\%) & 3793 (25.8\%)  & 66 (2.1\%) \\
        Fragmentation & 7567 (59.4\%) & 10928 (74.2\%) & 3117 (97.9\%) \\\hline
      \end{tabular}
    }
  \end{adjustwidth}
  \caption{Frequencies of two types of queueing delay.}
  \label{tab:queueing_delay}
\end{table}

\begin{figure*}[!t]
\begin{adjustwidth}{-1in}{-1in}
\centering
\subfigure[Passed]{\includegraphics[width=2.0in]{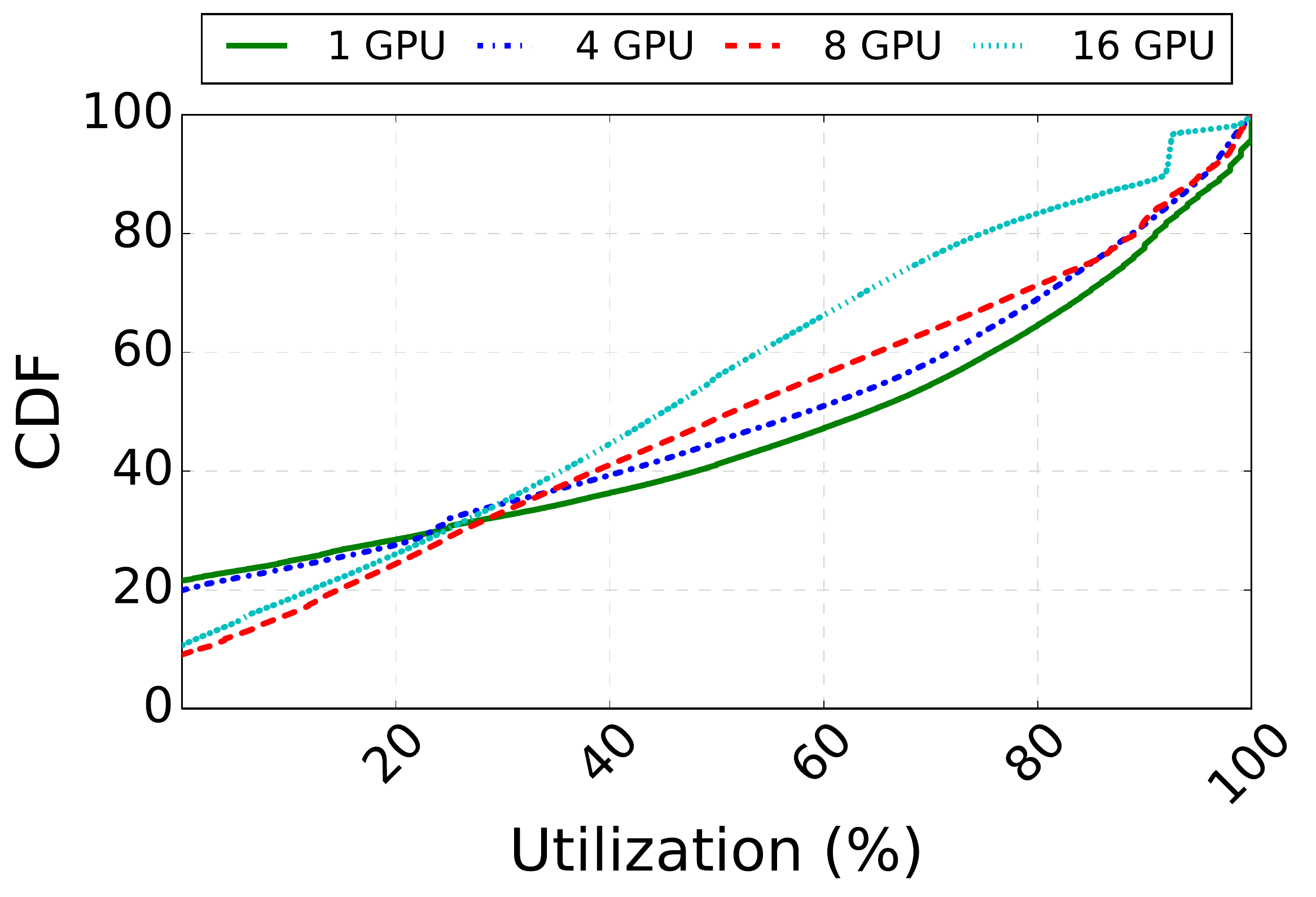}}
\subfigure[Killed]{\includegraphics[width=2.0in]{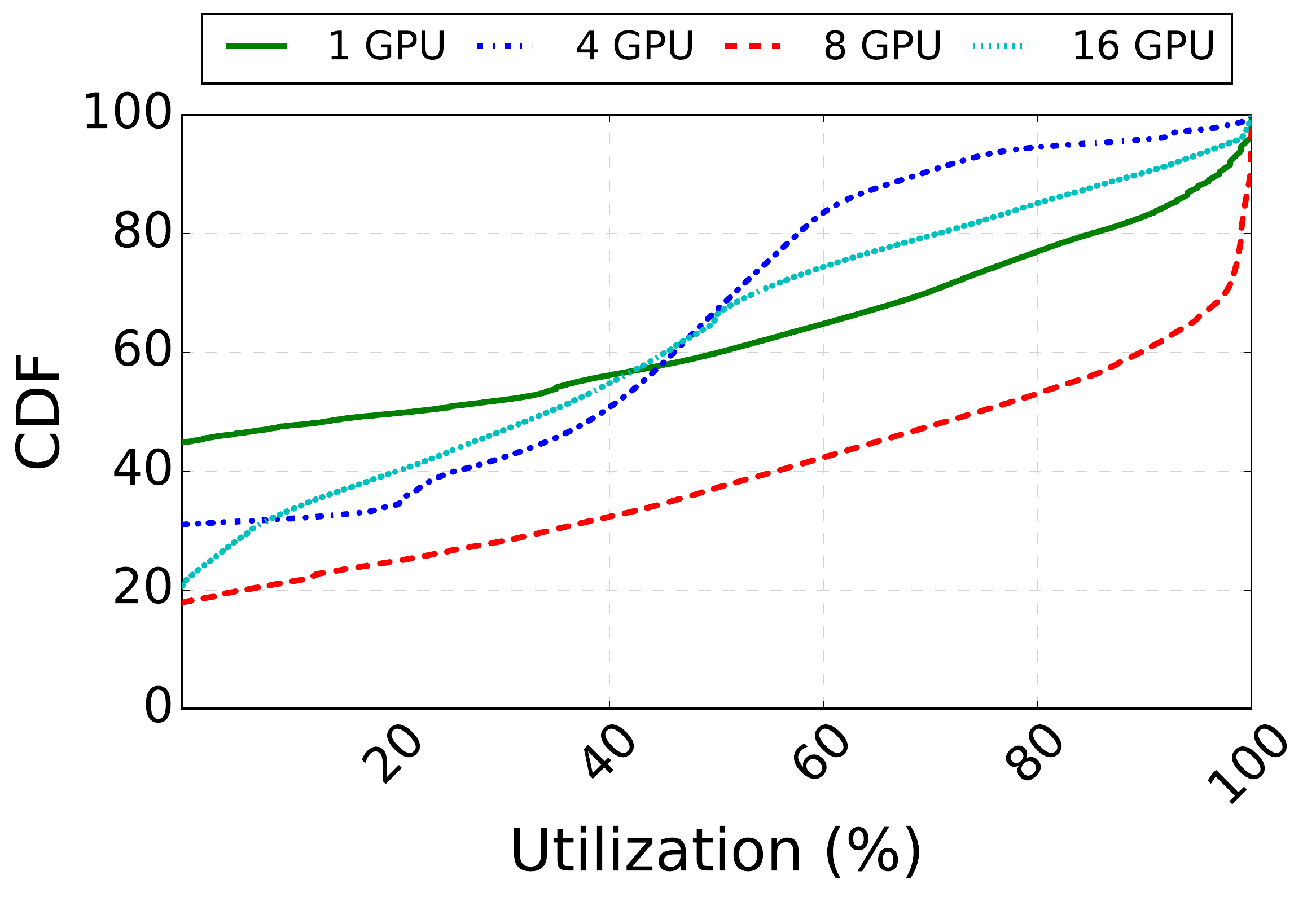}}
\subfigure[Unsuccessful]{\includegraphics[width=2.0in]{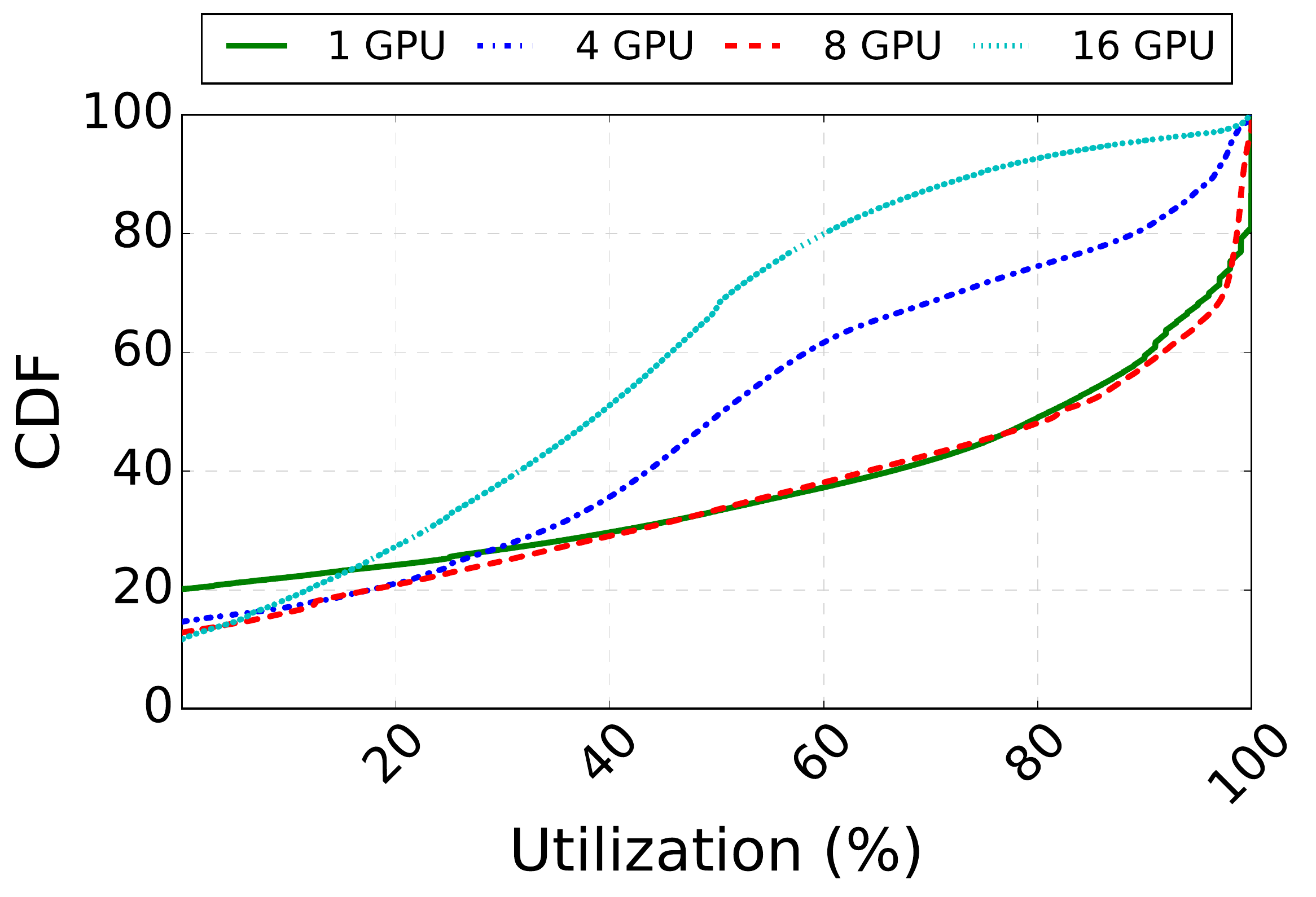}}
\end{adjustwidth}
\caption{\label{fig:gpu_dist} CDF of per-minute GPU utilization for passed, killed, unsuccessful jobs in different sizes. }
\end{figure*}


We next see how frequently fair-share delay and fragmentation delay occur for
different job sizes in our workloads. Since some jobs are quickly terminated,
we only consider jobs that run for at least one minute. \camera{Further,
since fragmentation influences distributed training jobs only, we consider
jobs that use 2 or more GPUs. Table~\ref{tab:queueing_delay} shows the
frequencies for the two types of delay. For jobs with 5-8 GPU, fragmentation
delay is responsible for 74.2\% of occurrences, and it dominates for larger
jobs. In contrast, for smaller jobs, we see that the two causes are more
balanced.} Further, we also observe that across all jobs fragmentation delay
is responsible for around 80\% of the delay in terms of waiting time. This is
because fair-share delays are easy to mitigate with preemption, but
fragmentation delays are much harder to overcome in our current design.


Finally, we note that the queuing delay fractions vary across virtual
clusters. Among the five largest virtual clusters, VC5 often over-subscribes
its quota and thus the proportion of fair-share delay is overall higher at
37\%.

\Paragraph{Does out-of-order scheduling exacerbate job queueing?} Given the
resource fragmentation and the fact that the YARN scheduler is
work-conserving, larger jobs could be additionally negatively affected by
out-of-order scheduling. To see how, consider a job that requires 24 GPUs
spread across three machines. While this job is waiting for such
configuration, if a smaller job requests 2 GPUs, it is scheduled on machines
where two GPUs become available. This could cause further fragmentation and
lead to the 24-GPU job needing to retry after a backoff. In our workload,
out-of-order scheduling is quite common, with 38.1\% of scheduling decisions,
and occurs 100\% for jobs with \camera{5-8 GPU or $>$8 GPU}. However, we find
that most out-of-order scheduling decisions do not greatly affect the waiting
time for resource-intensive jobs. For example, for out-of-order scheduling
occurrences of jobs with \camera{5-8 GPU or $>$8 GPU}, as much as 85.0\%
corresponds to cases where idle GPUs are effectively utilized without
prolonging the scheduling time of those waiting jobs.


In summary, our analysis shows why it makes sense to relax locality over time
to mitigate queuing delays for distributed training. We also find that in
addition to fair-share queuing delay, the need for gang scheduling and
locality introduces fragmentation delay for machine learning jobs.

\subsection{GPU utilization} \label{sec:GPU_utilization_analysis}


\label{sec:GPU_utilization} GPUs are the most expensive resources in our
cluster and this makes their efficiency an important factor in assessing the
cost-effectiveness across the entire cluster. For each individual GPU,
Ganglia~\cite{ganglia} reports aggregate performance counters every minute,
including utilization of processing cycles and memory, temperature, power
usage, etc~\cite{nvml}. We next present how efficiently training jobs use
processing cycles in their (exclusively) allocated GPUs. Note that our
current generation of GPUs only report coarse-grained utilization for
processing cycles that can only be used to detect if any of the streaming
multiprocessors (SMs) are being utilized~\cite{nvml}.  They do not report
what fraction of the SMs are being actually used within a single GPU.
Therefore, our analysis presents an ``upper bound'' of actual effective SM
utilization.


Overall, deep learning training jobs underutilize GPU processing cycles
regardless of their job sizes. Figure~\ref{fig:gpu_dist} shows CDFs of
per-minute GPU utilization of passed, killed, and unsuccessful jobs for
different sizes. Table~\ref{tab:avg_util} reports averages for each job size,
including averages for different job status; \camera{we use these job sizes
as representative of small, medium and large jobs based on the GPU request
distribution in our cluster}. Surprisingly we find that around 47.7\% of
in-use GPUs' cycles are wasted across all jobs, with jobs using 16 GPUs
exhibiting the lowest utilization at 40.39\%.
Moreover, across job status in Figure~\ref{fig:gpu_dist}, the median
utilization for 16 GPU jobs is 45.00\%, 34.24\%, 39.54\% for \texttt{Passed},
\texttt{Killed}, and \texttt{Unsuccessful}, respectively. These are 6.46\%,
40.25\%, and 42.63\% lower than the 8 GPU jobs in the corresponding job
status. We study the efficiency of such jobs in the next section in detail.

\begin{table}[!t]
\begin{adjustwidth}{-1in}{-1in}
\centering
\small
{
\begin{tabular}{c|ccc|l}
\hline
Job size & Passed & Killed & Unsuccessful & All \\\hline
1 GPU & 53.51 & 37.02 & 62.82 & 52.38 \\
4 GPU & 51.13 & 34.39 & 50.95 & 45.18 \\
8 GPU & 51.09 & 60.63 & 64.34 & 58.99 \\
16 GPU & 44.88 & 36.98 & 39.02 & 40.39 \\\hline
All & 52.43 & 42.98 & 60.43 & 52.32 \\\hline
\end{tabular}
}
\end{adjustwidth}
\caption{Mean GPU utilization for different job sizes.}
\label{tab:avg_util}
\end{table}

\subsubsection{Impact of Distributed Learning} \label{sec:distribution_impact}

Given that the 8 GPUs mounted in each server can communicate more efficiently
without using the network, our job scheduling strategy is to favor
intra-server locality when assigning each job to available GPUs. At the same
time, the scheduler attempts to pack small jobs into fewer servers to avoid
fragmentation. This leads to \emph{job colocation} on the same server and
consequently could lead to interference in shared system resources (e.g.,
RDMA and PCIe)~\cite{wencong1}. This creates an interesting utilization
spectrum for multi-GPU jobs. In particular, jobs using more than 8 GPUs must
\emph{distribute} training instances across multiple servers and may be
dynamically colocated with other jobs. This scenario also involves
communication overheads since each server has to periodically wait for model
aggregation to happen over the network.

To confirm that such distribution and colocation factors indeed relate to the
efficiency of GPUs in use, we first characterize utilization of processing
cycles for various job placement scenarios using a popular image recognition
model, ResNet-50~\cite{he2016deep}. Specifically we train ResNet-50 with 2
GPUs using TensorFlow and perform offline experiments with
placements that exercise shared resources differently. Then using our
telemetry data, we attempt to infer correlations between those factors and
the observed efficiency in our cluster.

\begin{table}[!t]
\begin{adjustwidth}{-1in}{-1in}
\centering
\small
{
\begin{tabular}{c|cc|cc}
\hline
Metric & SameServer & DiffServer & IntraServer & InterServer \\\hline
GPU util. & 57.7 & 49.6 & 37.5 & 36.5 \\
Images/s & 114.8 & 98.0 & 75.6 & 74.1 \\\hline
\end{tabular}
}
\end{adjustwidth}
\caption{Mean GPU utilization and training performance of ResNet-50 over different locality/colocation configurations.}
\label{tab:resnet}
\end{table}

\Paragraph{Analysis using ResNet-50.} Table~\ref{tab:resnet} shows the impact
of distribution only, by comparing a ResNet-50 job placed in a single server
(\texttt{SameServer}) with the job placed in two servers connected with RDMA
network (\texttt{DiffServer}). Each server has four NVIDIA Tesla P100 GPUs
attached to a CPU socket. The table reports GPU utilization when processing a
batch size of 32 images during training. First we observe that the training
does not fully utilize GPUs even for single machine execution. In particular,
\texttt{SameServer} achieves utilization of 57.7\% for GPUs in use. It
increases to 71.1\% for twice the batch size but only increases marginally
for larger batches. Also the table shows that using distributed training
achieves lower utilization of 49.6\% in \texttt{DiffServer}. This shows that
even for 2-GPU jobs, there is a cost to not achieving locality.

\begin{figure}[!t]
  \centering
  \vspace{-\intextsep}
  \hspace*{-.75\columnsep}
  \includegraphics[width=2.1in]{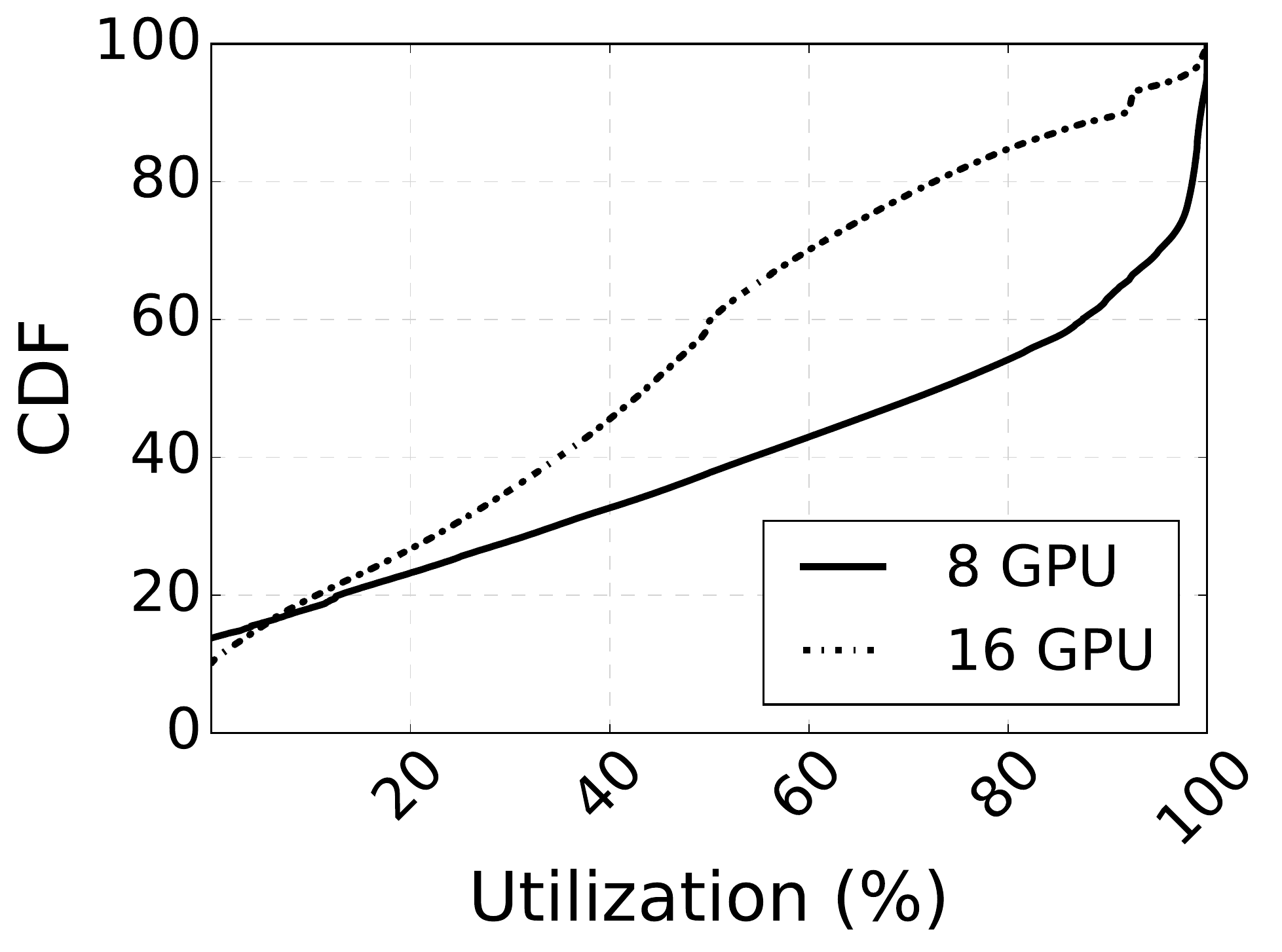}
  \caption{\label{fig:dist_mode} GPU utilization when running 8 and 16 GPU jobs on dedicated servers.}
  \vspace{-0.1in}
\end{figure}

Given a distributed training setup, contention for shared resources like
RDMA and PCIe further lowers the efficiency of utilized GPUs. To show this we
set \texttt{DiffServer} as our baseline and measure changes in the efficiency
while populating additional ResNet-50 jobs in the same servers. First, we
measure GPU utilization when the colocated jobs do not
use RDMA network at all: we place two \texttt{SameServer} jobs, one on each server
in the same CPU socket as the job under study.
Thus, these jobs interfere with the
job under study in the use of PCIe buses while reading training inputs, aggregating
model updates, and so on. The observed efficiency is shown as
\texttt{IntraServer} in Table~\ref{tab:resnet}, and we see that having such
intra-server interference lowers the utilization by as much as
12.1\%. We also study if such interference matters for the RDMA network
in \texttt{InterServer}. For this setup we use two \texttt{DiffServer} jobs instead of
two \texttt{SameServer} jobs as background traffic, so that all the jobs
are distributed across two servers and share the RDMA network. In this case, we see
a 13.1\% decrease in utilization compared to the baseline.

Our experimental study reveals that efficiency of allocated GPUs varies
according to locality and colocation scenarios that could occur in the
cluster. Further, any placement that causes lowered GPU utilization also
results in slowdown in training performance (i.e., images processed per
second) as shown in Table~\ref{tab:resnet}. Next, we analyze utilization for
our aggregate workload. We note that unlike the controlled experiment,
the type of model trained and the batch sizes used vary across jobs in our
aggregate workload making it harder to establish a baseline utilization
without distribution or inference.

\Paragraph{Distributed training with dedicated servers.} First, to study the effects of distribution,
we restrict our study to look at 8 GPU and 16 GPU jobs that are packed on
one or two servers. In this case, the 8 GPU jobs uses all 8 GPUs in a single server
while the 16 GPU jobs uses all the GPUs in two servers.
The network over which the servers for these jobs are connected to each other is shared.
Figure~\ref{fig:dist_mode} shows the results of our comparison. Compared to
the 8 GPU jobs, we see that 16 GPU jobs, which have the additional model
aggregation step in distributed mode, have significantly lower utilization.
Specifically, for 8 GPU jobs, GPU cycles are utilized 56.9\% of time on
average while this is only 34.3\% for 16 GPU jobs. Furthermore, the median is
73.12\% for 8 GPU jobs, which is 1.67x the median in the 16 GPU case.

\Paragraph{Distributed training with shared servers.} When locality
constraints are relaxed, a job may have to be distributed over many servers
while sharing them with other jobs. Distributing a job over many shared
servers can further lower utilization of GPUs. This drop in utilization
occurs not only due to a higher network overhead but also because of
interference from unrelated but co-located jobs. To study this, we see how
the GPU utilization of 16-GPU jobs varies as as we move from dedicated GPUs
to a larger number of shared servers. Table~\ref{tab:dist_degree} shows the
average and percentiles for GPU utilization across the different allocation
scenarios.

When running on 2 8-GPU servers, a 16-GPU job has dedicated servers.
When running on 4 servers, the 16-GPU job may occupy 4 GPUs on each
server, and will be colocated with other jobs on those servers. We find that
the degree of interference is larger if the job is distributed on more
servers. Table~\ref{tab:dist_degree} shows that in addition to the
inefficiency caused by distribution (Figure~\ref{fig:dist_mode}) there is
additional underutilization caused by colocation. We see that for 16-GPU jobs
distributed across 8 servers, the average utilization is as low as 28.26\%
and more than 90\% of jobs have less than 66\% utilization.


\begin{figure}[!t]
  \centering
  \includegraphics[width=1.85in]{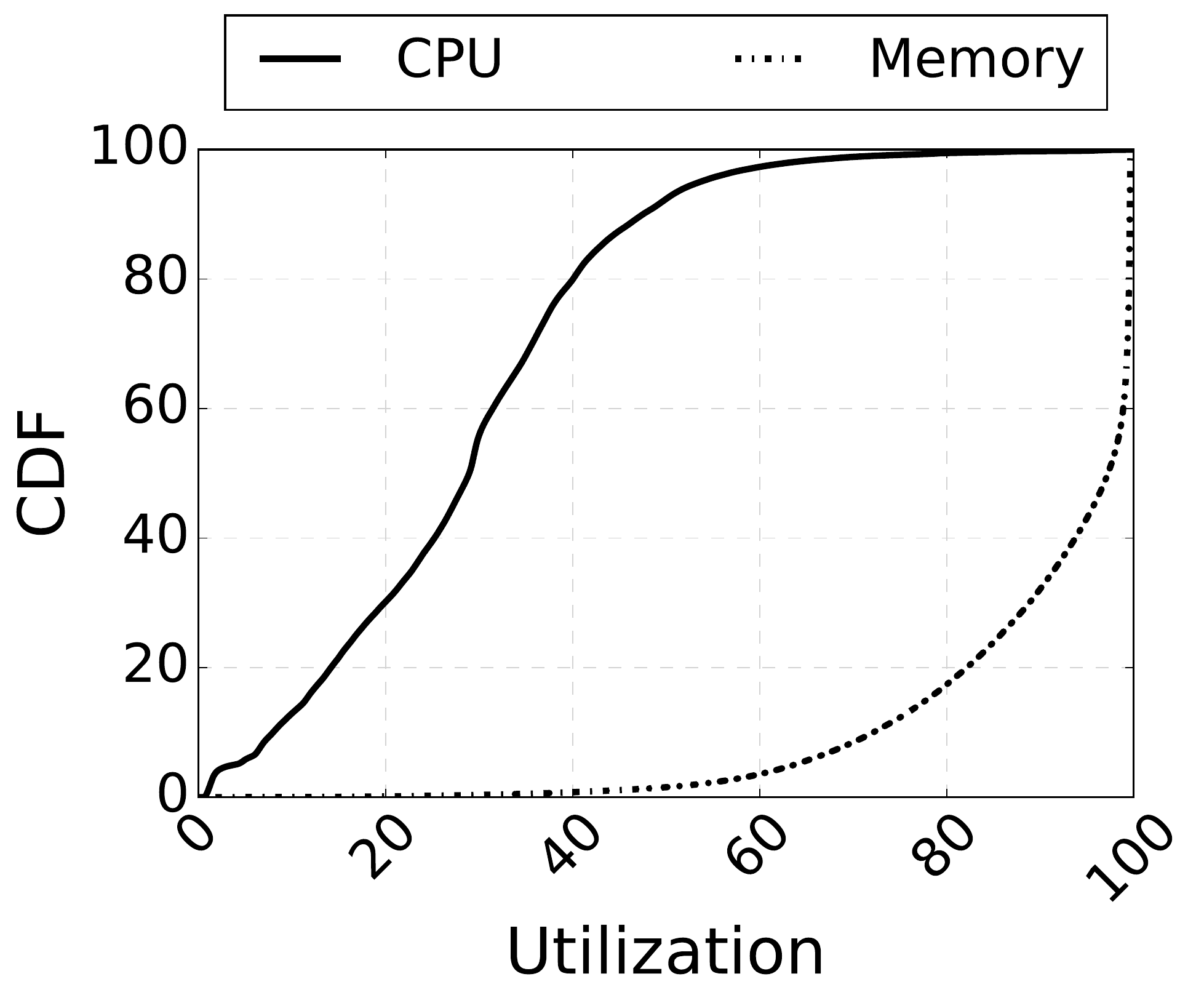}
  \vspace{-0.1in}
  \caption{\label{fig:dist_host} Host resource utilization.}
\end{figure}

\begin{table}[!t]
\begin{adjustwidth}{-1in}{-1in}
\centering
\small
{
\begin{tabular}{c|lllll}
\hline
Degree & Mean & 50\%ile & 90\%ile & 95\%ile \\\hline
2 servers & 43.66 & 43.69 & 91.77 & 97.06\\
4 servers & 40.94 & 39.85 & 83.28 & 91.97\\
8 servers & 28.56 & 25.71 & 65.68 & 78.85\\\hline
\end{tabular}
}
\end{adjustwidth}
\caption{GPU utilization for 16-GPU jobs that are spread over 2, 4, and 8 servers.}
\label{tab:dist_degree}
\end{table}



Among host resources, our scheduler dedicates CPU and memory along with GPU
to each job. In deep learning clusters, these host resources are used for
many useful tasks including caching training inputs, model aggregation, and
periodic model validation and progress report. By default, we allocate CPU
and memory capacity proportional to the number of requested GPUs.
Figure~\ref{fig:dist_host} shows CDFs of utilization of these host resources
observed in our servers. In general, many servers underutilize CPU cycles yet
highly utilize memory. This indicates that a useful feature in the scheduler
would be to observe if a particular job requires disproportionate amount of
host memory and isolate memory used by jobs colocated on the same server.

In summary, our data analysis shows how GPUs are underutilized in shared
clusters. We presented correlations of how distribution and interference
affect utilization and validated this using a controlled experiment to break
down the importance of locality and interference. We
discuss some implications for scheduler design in
Section~\ref{sec:Implications}.

\section{Training Progress and Completion}

Jobs in our system finish with one of three statuses: passed, killed or
unsuccessful. Similar to iterative online
computations~\cite{Condie10,Agarwal13}, our machine learning job utilizes
cluster resources to improve the model over time.
However as opposed to prior study on big data traces~\cite{Kavulya10}, we see
a significant fraction of jobs (30.7\% as shown in
Table~\ref{tab:pass_killed_unsucc}) are either terminated unsuccessfully or
killed by users. They constitute around 55\% of the total GPU time used
during our trace collection period. Thus it is important to understand the
reason behind these failures as fewer unsuccessful jobs would mean that more
of the cluster resources can be used for successful jobs.

\begin{table}[!t]
  \begin{adjustwidth}{-1in}{-1in}
    \centering
    \small
    {
      \begin{tabular}{c||c|c}
        \hline
        Status & Count(\%) & GPU times used (\%) \\\hline
        Passed & 66696 (69.3\%) & 44.53\% \\
        Killed & 12996 (13.5\%) & 37.69\% \\
        Unsuccessful & 16568 (17.2\%) & 17.76\% \\\hline
        Total & 96260 (100.0\%) & 100.0\%\\\hline
      \end{tabular}
    }
  \end{adjustwidth}
  \caption{Distribution of jobs by their final status.}
  \label{tab:pass_killed_unsucc}
  \vspace{-10pt}
\end{table}

\subsection{Effectiveness of Training Iterations}
\label{sec:ExcessiveIterations}


Most deep learning jobs optimize a non-convex loss function and the
optimization algorithms do not necessarily guarantee that the loss always
decreases with more training. Thus, similar to~\cite{Juncheng19}, users in
our system submit model training jobs using a larger number of epochs than
necessary to get the optimal model. To analyze the magnitude of this effect
we study how the training loss for a job varies across epochs and measure the
epoch at which we achieve the best training loss. As this information is not
printed in the log by every user/framework, we are only able to obtain
convergence information for around 2502 jobs.

\begin{figure}[!t]
\begin{adjustwidth}{-1in}{-1in}
\centering
\subfigure[Passed jobs]{\includegraphics[width=1.6in]{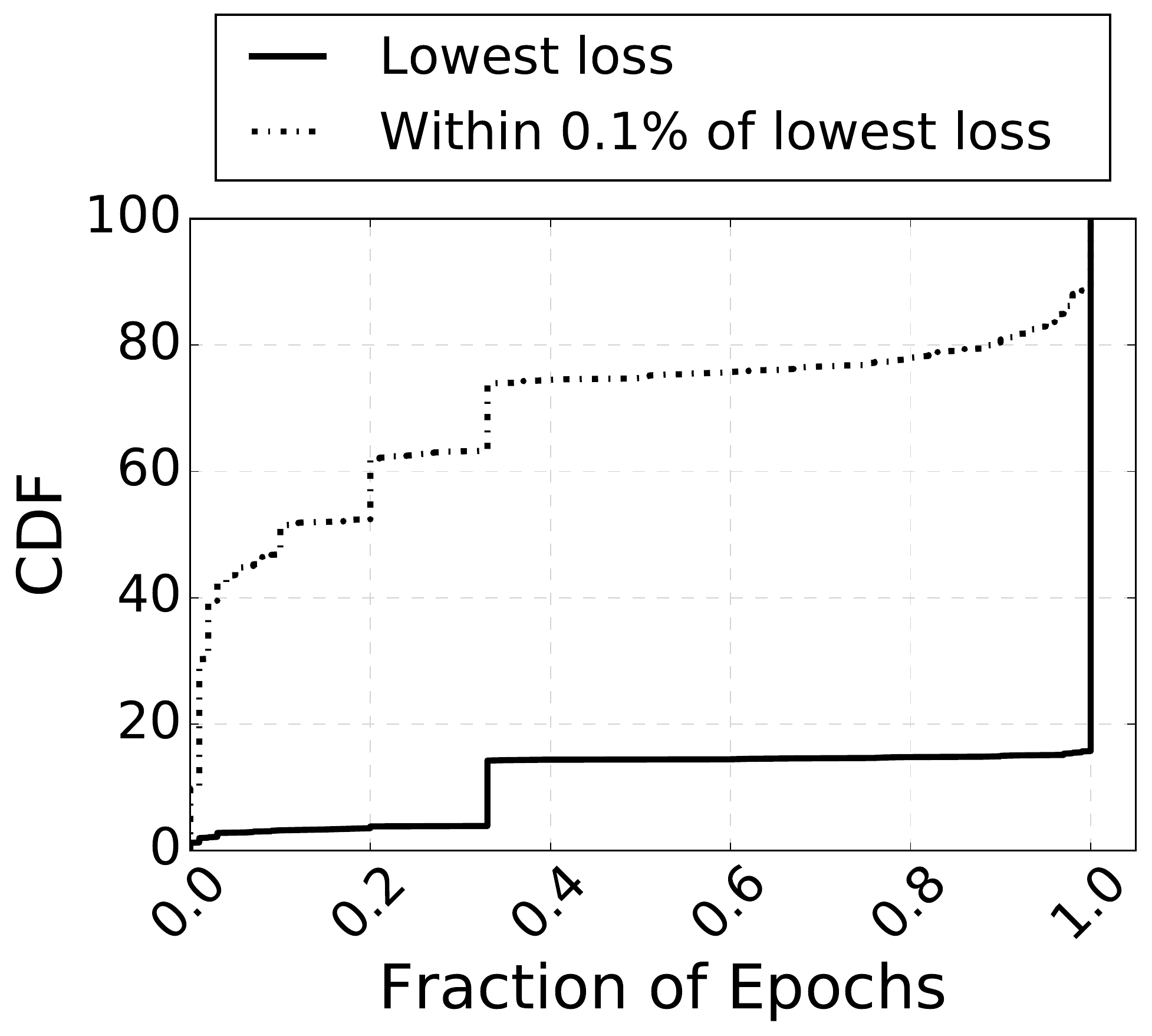}}
\subfigure[Killed jobs]{\includegraphics[width=1.6in]{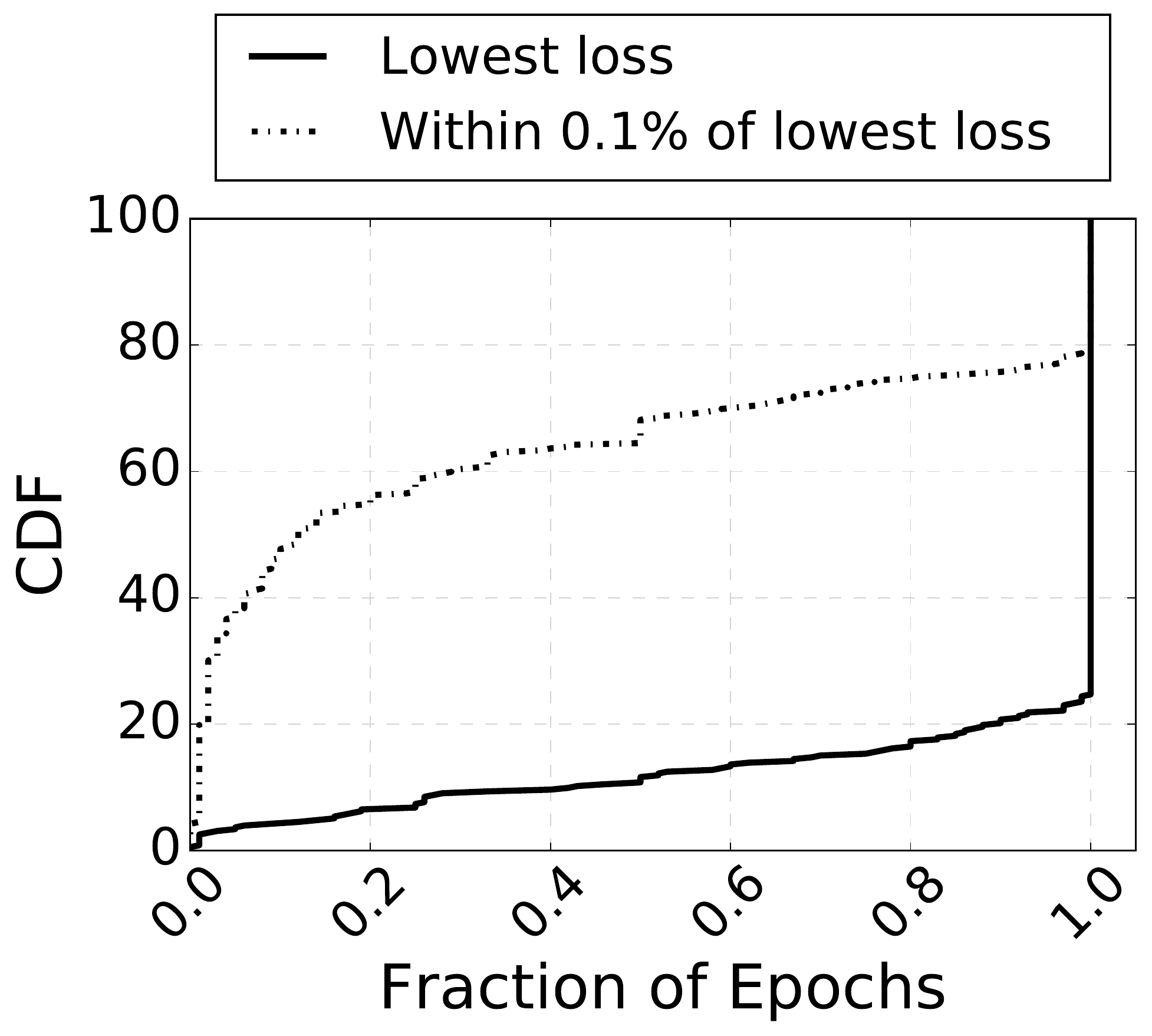}}
\end{adjustwidth}
\caption{\label{fig:train_loss} Fraction of epochs necessary to achieve a particular loss
  threshold for (a) passed jobs and (b) killed jobs.}
\end{figure}

First, Figure~\ref{fig:train_loss}(a) shows the fractions of epochs required
to reach the lowest loss across all passed jobs. From the figure we see that
around 80\% of passed jobs require all the epochs executed to reach the
lowest loss. We repeat this study for killed jobs and see a similar pattern
as shown in Figure~\ref{fig:train_loss}(b).

However we also see that a majority of jobs improve the loss marginally using
a large fraction of epochs. In particular, Figure~\ref{fig:train_loss}(a)
shows the fraction of epochs required to reach within 0.1\% of the lowest
loss across all passed jobs. Around 75\% of jobs reach within 0.1\% of the
lowest loss using only 40\% of the epochs. Again, a similar pattern is shown
for killed jobs in Figure~\ref{fig:train_loss}(b). While we do not present
data from user surveys, this suggests that machine learning practitioners can
early terminate jobs to save use of GPU times considerably when the loss
change is less than a particular threshold in successive epochs. Essentially,
we look into how much resources are used to improve 0.1\% of convergence
accuracy in terms of the fraction of GPU times for each job. In our workload,
this accounts for 62\% and 56\% on average for passed jobs and killed jobs,
respectively.


%

\subsection{Job Failures}
\label{sec:JobFailures}

\begin{figure}[!t]
\begin{adjustwidth}{-1in}{-1in}
\centering
\subfigure[Retries]{\includegraphics[width=1.7in]{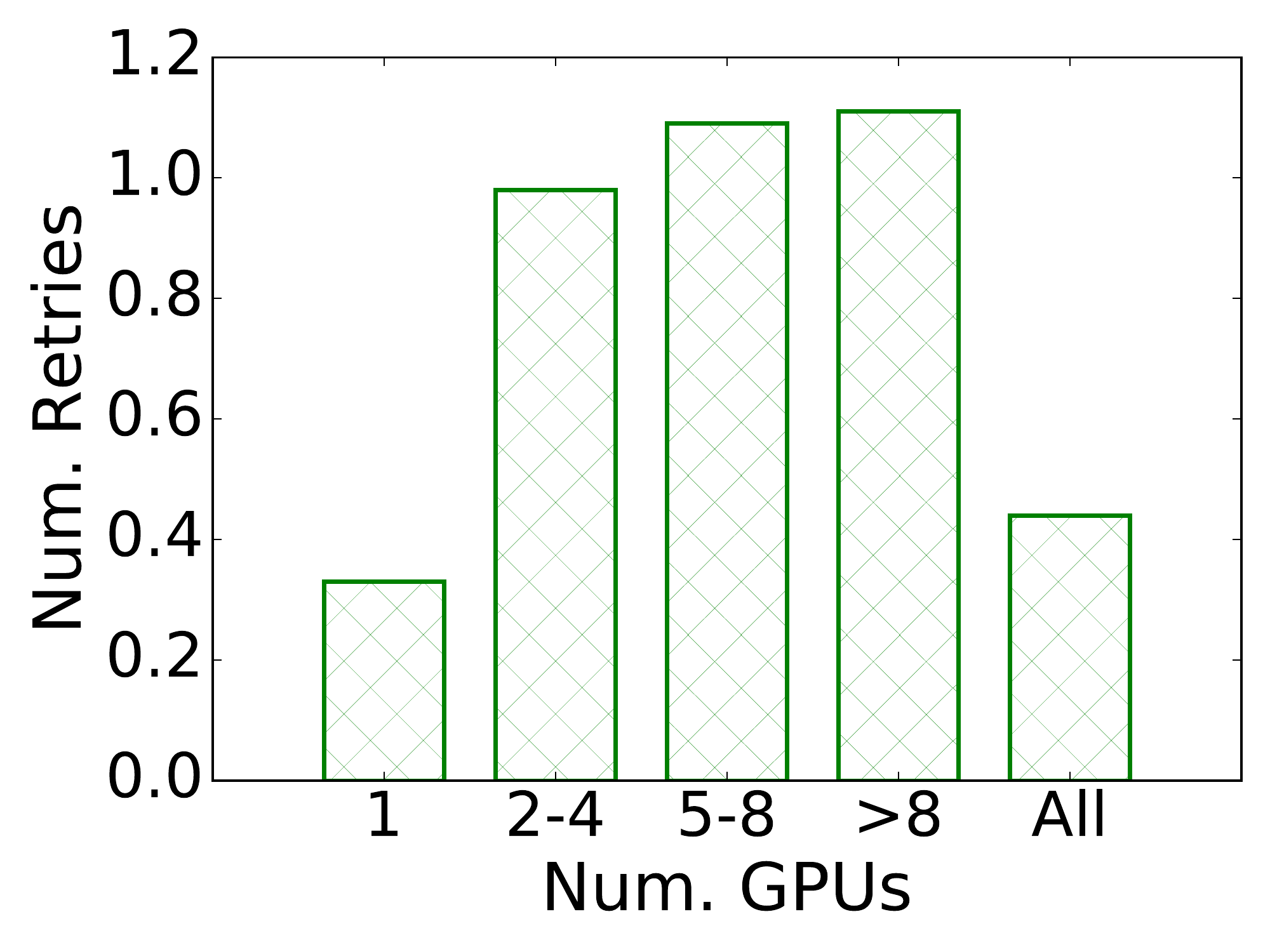}}
\subfigure[Unsuccessful jobs]{\includegraphics[width=1.7in]{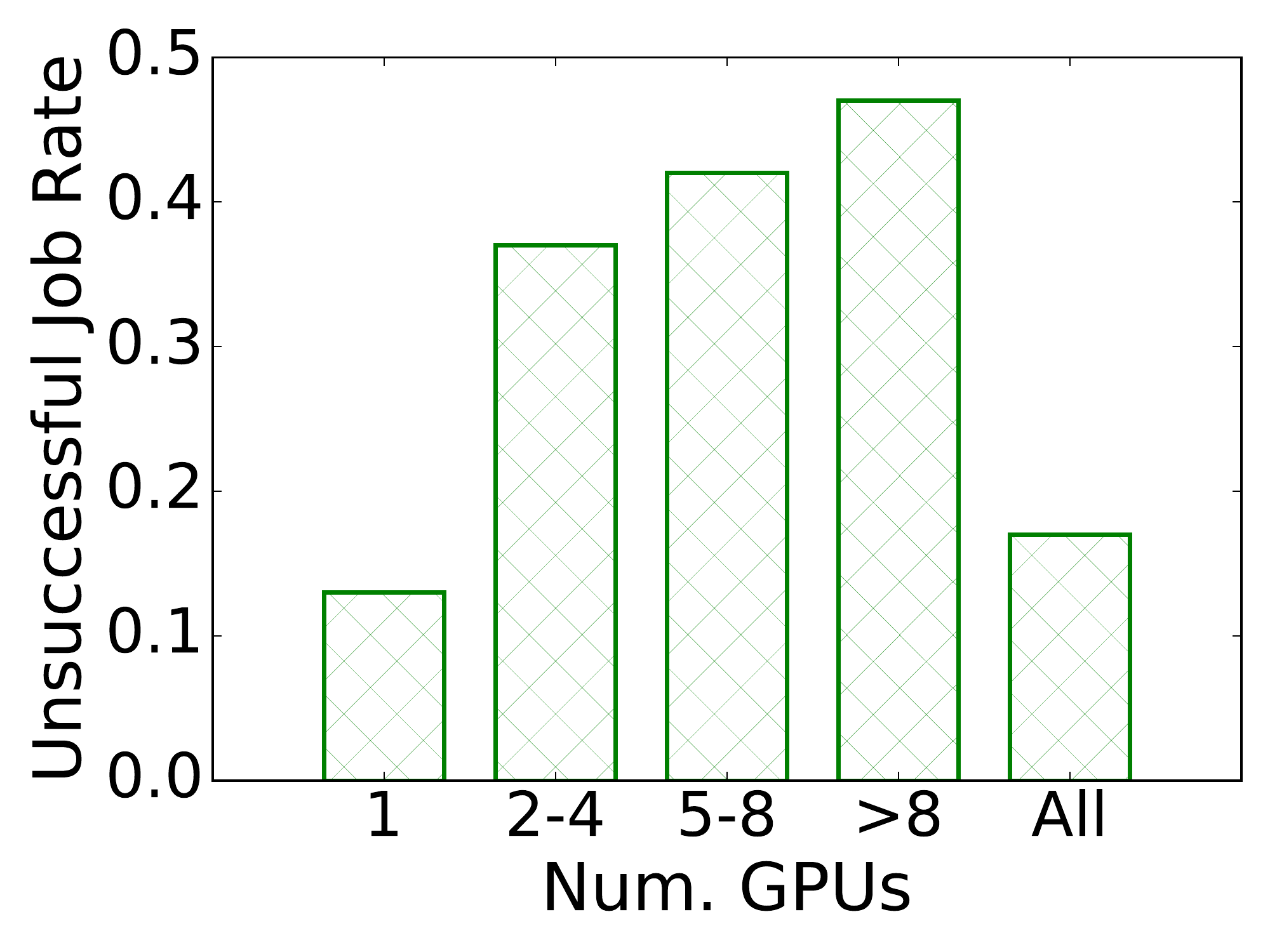}}
\end{adjustwidth}
\caption{\label{fig:gpu_retries_failure} (a) Average number of job retries for using different number of GPUs, and (b) subsequent unsuccessful jobs.}
\end{figure}

We next present a detailed analysis on job failures, including why/when/how
frequently jobs fail and what their impact is on effective cluster usage. We
remind the reader that in our cluster scheduler, a job is retried upon
failure. If the job repeatedly fails it is marked as unsuccessful as further
retries are deemed no longer effective. Figure~\ref{fig:gpu_retries_failure}
presents a high-level summary of job retries/failures and shows that jobs
using \camera{more than 4} GPUs not only retry execution more often but also
finish unsuccessfully at higher rate.  The reasons behind job
retries/failures are diverse, and failures occur at different times during
job execution. We thus investigate failures by classifying them across layers
of our system stack.

\begin{table*}[!t]
\begin{adjustwidth}{-2in}{-2in}
  \setlength\tabcolsep{4.4pt} 
\centering
\footnotesize
  \begin{tabular}{p{2.4cm}||ccc|ccc|rrr|r|ccc|p{2cm}}
\hline
\multirow{2}{*}{{\bf Failure Reason}} &
\multicolumn{3}{c|}{\bf Category} &
\multicolumn{3}{c|}{\bf Num Occurrences} &
\multicolumn{4}{c|}{\bf RTF: Runtime to Failure (mins)} &
\multicolumn{3}{c|}{\bf GPU Demand} &
\multirow{2}{*}{\shortstack{{\bf RTF$\times$Demand (\%)}}}\\
& IF & AE & U & Trial & Job & User & 50\%ile & 90\%ile & 95\%ile & Total \% & 1 & 2-4 & $>$4 &\\

\hline\hline
CPU out of memory & & \cmark & \cmark & 12076 & 2803 & 65 & 13.45 & 17.73 & 33.97 & 6.62 & 11465 & 235 & 376 & 3982320 (8.05)\\
Incorrect inputs & \cmark & & \cmark & 9690 & 4936 & 208 & 1.87 & 404.83 & 2095.73 & 30.43 & 5844 & 2638 & 1208 & 11979474 (24.21)\\
Semantic error & \cmark & & \cmark & 2943 & 2049 & 159 & 2.72 & 376.00 & 1436.88 & 9.22 & 1603 & 494 & 846 & 8442835 (17.06)\\
Core dump & & \cmark & \cmark & 2912 & 1784 & 122 & 0.85 & 72.75 & 431.65 & 3.35 & 1936 & 496 & 480 & 1493632 (3.02)\\
Invalid mem access & & & \cmark & 2602 & 1235 & 108 & 1.03 & 403.50 & 1357.38 & 3.82 & 712 & 774 & 1116 & 2352994 (4.75)\\
Model ckpt error & \cmark & & & 1995 & 948 & 85 & 181.67 & 3728.93 & 8196.02 & 21.73 & 743 & 384 & 868 & 8080374 (16.33)\\
CUDA failure & & \cmark & & 1484 & 571 & 70 & 1.32 & 19.87 & 82.17 & 0.62 & 133 & 1153 & 198 & 357119 (0.72)\\
Syntax error & \cmark & & \cmark & 1132 & 883 & 110 & 0.58 & 5.02 & 12.00 & 0.19 & 780 & 184 & 168 & 130094 (0.26)\\
Traceback from crash & \cmark & \cmark & \cmark & 777 & 271 & 44 & 1.02 & 894.33 & 1394.07 & 2.34 & 356 & 277 & 144 & 863130 (1.74)\\
MPI error & \cmark & & & 634 & 166 & 28 & 1.62 & 3015.27 & 5143.98 & 3.70 & 456 & 54 & 124 & 613059 (1.24)\\
GPU out of memory & & \cmark & & 487 & 261 & 35 & 18.53 & 353.62 & 2740.28 & 1.08 & 237 & 70 & 180 & 1040249 (2.10)\\
MPI runtime failure & \cmark & & & 478 & 420 & 96 & 1389.48 & 13778.60 & 18090.88 & 14.63 & 240 & 141 & 97 & 7593398 (15.34)\\
Permission error & & & \cmark & 299 & 151 & 37 & 1.00 & 8.15 & 15.85 & 0.07 & 56 & 202 & 41 & 15185 (0.03)\\
Import error & \cmark & & \cmark & 148 & 148 & 41 & 0.67 & 4.58 & 10.73 & 0.06 & 108 & 30 & 10 & 10803 (0.02)\\
Job preempted & \cmark & & & 147 & 95 & 34 & 559.08 & 2682.85 & 5892.23 & 1.66 & 25 & 95 & 27 & 2338772 (4.73)\\
CUDA init failed & & \cmark & & 141 & 69 & 20 & 1.08 & 2.18 & 4.63 & 0.03 & 16 & 66 & 59 & 64512 (0.13)\\
Model diverged & & & \cmark & 84 & 30 & 5 & 1.48 & 44.37 & 76.53 & 0.01 & 78 & 5 & 1 & 2562 (0.01)\\
CUDA ver. mismatch & & \cmark & & 49 & 49 & 19 & 0.83 & 1.65 & 1.67 & 0.00 & 1 & 1 & 47 & 421 (0.00)\\
GPU ECC error & & \cmark & & 10 & 10 & 2 & 26.82 & 671.92 & 2035.02 & 0.03 & 1 & 5 & 4 & 23575 (0.05)\\
Output node error & & & \cmark & 3 & 3 & 1 & 0.85 & 0.95 & 0.95 & 0.00 & 3 & 0 & 0 & 2 (0.00)\\
Cannot load libs & & \cmark & & 1 & 1 & 1 & 0.12 & 0.12 & 0.12 & 0.00 & 1 & 0 & 0 & 0.12 (0.00)\\
\hline
No signature & & & & 1684 & 698 & 94 & 1.87 & 28.00 & 95.17 & 0.42 & 1235 & 294 & 155 & 102138.03 (0.21)\\
\hline

\end{tabular}
\end{adjustwidth}
\caption{Failures classified into failure reasons (sorted based on the number of occurrences). There are
  largely three categories that cause the failures: Infrastructure (IF), AI Engine (AE), and
  User (U). A failure reason may be observed in multiple categories.}
\label{table:failures}
\end{table*}

\subsubsection{Failure Classification}
Table~\ref{table:failures} presents analysis results of failures based on two
classification factors. First, failures are classified from different sources
(Column 2): the sources include (i) Infrastructure (\texttt{IF}) which
includes YARN, HDFS and all other framework components, (ii) AI Engine
(\texttt{AE}) which includes TensorFlow, Torch, and any other platforms, and
(iii) User (\texttt{U}) which represents programmers. Column 1 lists a number
of reasons for failures we observe from the workload.

Most failure reasons in the table are self-explanatory, and we describe six
important ones in more detail here.
\begin{myenumerate}
\item[(1)] \textbf{Incorrect inputs}: Model files or input data
    stored in the external HDFS storage cannot be read.
\item[(2)] \textbf{Semantic error:} Errors that happen due to library version
    mismatch or other dependencies of the user training program not being
    setup correctly.
\item[(3)] \textbf{Model checkpoint error:} The job is not able to
    successfully create a model checkpoint after a certain number of epochs
    complete. This is usually due to either transient error in HDFS or HDFS
    name node recovery.
\item[(4)] \textbf{MPI runtime failure:} This is usually due to either a
    failure of network connection to peer MPI process, or possibly an internal
    failure of the MPI daemon itself.
\item[(5)] \textbf{Job preempted:} YARN reclaims any GPU currently in use to
    schedule another job.
\item[(6)] \textbf{Invalid memory access:} Training job dies because of
    violating access on memory address space, e.g., using an invalid pointer
    value, or having race condition while copying data. This failure is
    observed in both CPU memory and memory allocated for GPU access.
\end{myenumerate}

\noindent While bridging failure category and failure reason, we observe that
a failure reason can appear in multiple categories, even in all involved
categories, as shown in Column 2 of Table~\ref{table:failures}.

\Paragraph{Building failure classifier.} There exists causality among various
failure reasons. For example, \emph{traceback from crash} is a consequence
of an \emph{invalid memory access}.  Our first mission in building a
classifier is identifying signatures of failure reasons closer to the root
cause.
We capture root-cause signatures from stdout or stderr logs of a failed job.
If not explicit from the logs, we then attempt to capture implicit ones such
as \emph{traceback from crash}. In consequence, our classifier has in total
\textbf{more than 230 rules} to find both explicit signatures and implicit
signatures. If there is no signature for a failure, we tag it as
\emph{no signature}, which constitutes 4.2\% of the total
failures.

\subsubsection{Failure Frequency}

Column 3 of Table~\ref{table:failures} summarizes the occurrence frequency of
the classified failure reason. \texttt{Trial} counts the number of failure
events observed in our workload: failure reasons are sorted by it. We further
group \texttt{Trial} occurrences by job ID (\texttt{Job}) and user ID
(\texttt{User}) to see if failures are localized according to the same job or
user.

\Paragraph{Failures repeat for the same job/user.} Our analysis shows that
across failure reasons, failures repeat at both job level and user level. In
particular, we measure repetition factors (i.e., \texttt{Trial} divided by
\texttt{Job} or \texttt{User}) for top-8 failure reasons, which cover 88.9\%
of the total failures. The measured repetition factors are 2.3 and 38.8 on
average for \texttt{Job} and \texttt{User}, respectively, meaning a single
job and a single user on average cause 2.3 and 38.8 occurrences of failure,
respectively, during the data collection period. The most critical one is
\emph{CPU out of memory}, where we see 185.7 as the \texttt{User} repetition
factor. Interestingly, profiling shows that a certain engineer issued a
number of training jobs, all of which suffer from the same out-of-memory
issue, resulting in high concentration of failures. This motivates the need
for runtime detection mechanisms that can correlate errors from the same user
even though her jobs are independent from job management point of view.

\Paragraph{User/programming errors lead to a lot of failures.} Failures
incurred by user errors, such as configuration/syntax/semantic errors in
program and script, are dominant. These failures are very prevalent across
our top-8 failure reasons. As explained, \emph{CPU out of memory} is the most
frequent with its failures significantly concentrated on a few users. Other
frequent failures such as \emph{incorrect inputs} and \emph{semantic error}
are more spread out across different users. From our profiling, the primary
factor that causes those failures is a lot of independent components involved
in a training job. For example, by definition, \emph{incorrect inputs}
happens when there is a failure in reading model or input data stored in
external HDFS store. This is due to any error along the path of accessing
data from user program/script to data stored in the HDFS: the path is not
correct, data format is inconsistent, data itself is corrupted, etc.
Often, issues in data format affect multiple engineers in the same team
(e.g., speech recognition team) as they often share the
same training data or reference model. 

\enlargethispage{\baselineskip}

\subsubsection{Runtime to Failure}
Column 4 of Table~\ref{table:failures} presents runtime to failure (RTF) for
each classified failure reason. To capture the summary of RTF distribution,
in addition to the average, we also present the 50th-percentile (or median)
and higher percentiles such as 90th-percentile and 95th-percentile.

\Paragraph{The runtime to failure (RTF) exhibits high variability, with
mainly short RTFs.} Many failures of training jobs happen quickly, for
example within 10 mins. This is mostly the case for failures driven by users
in syntax, semantic, and configuration errors, which we can also infer from
low median RTFs in the corresponding failure reasons. Note that most of those
failures are deterministic and are caught when the runtime begins to execute
the program. One of exceptions that is noteworthy is failure corresponding to
inconsistent/corrupted input data. We can only detect this at the moment we
actually read the erroneous data and attempt to parse it. This is the primary
reason for having high 95th-percentile in \emph{incorrect inputs}.

\Paragraph{Infrastructure failures occur infrequently but have much longer
runtime to failure (RTF).} This analysis focuses on two failure reasons in
infrastructure category: \emph{model checkpoint error} and \emph{MPI runtime
failure}. They represent program-to-storage and program-to-program
communication, which are both critical for reliable distributed deep learning
training. In general, these errors are relatively infrequent compared to
other common errors, constituting only 6.2\% of the total \texttt{Trial}s.
However, our analysis shows that these errors tend to appear after a long
execution duration, and thus dominate the time until failure detection.  In
particular, Table~\ref{table:failures} shows that when the corresponding RTFs
are summed up (i.e. \texttt{Total}), the two failure reasons, \emph{model
checkpoint error} and \emph{MPI runtime error}, occupy as much as 21.73\% and
14.63\%, respectively.

\subsubsection{Impact of GPUs Allocated}

For jobs with the same RTF, the impact on the amount of utilized resources is
proportional to the number of allocated GPUs. The larger
the allocation, the bigger the impact.

\begin{figure}[!t]
\begin{adjustwidth}{-1in}{-1in}
\centering
\subfigure[Incorrect inputs]{\includegraphics[width=1.7in]{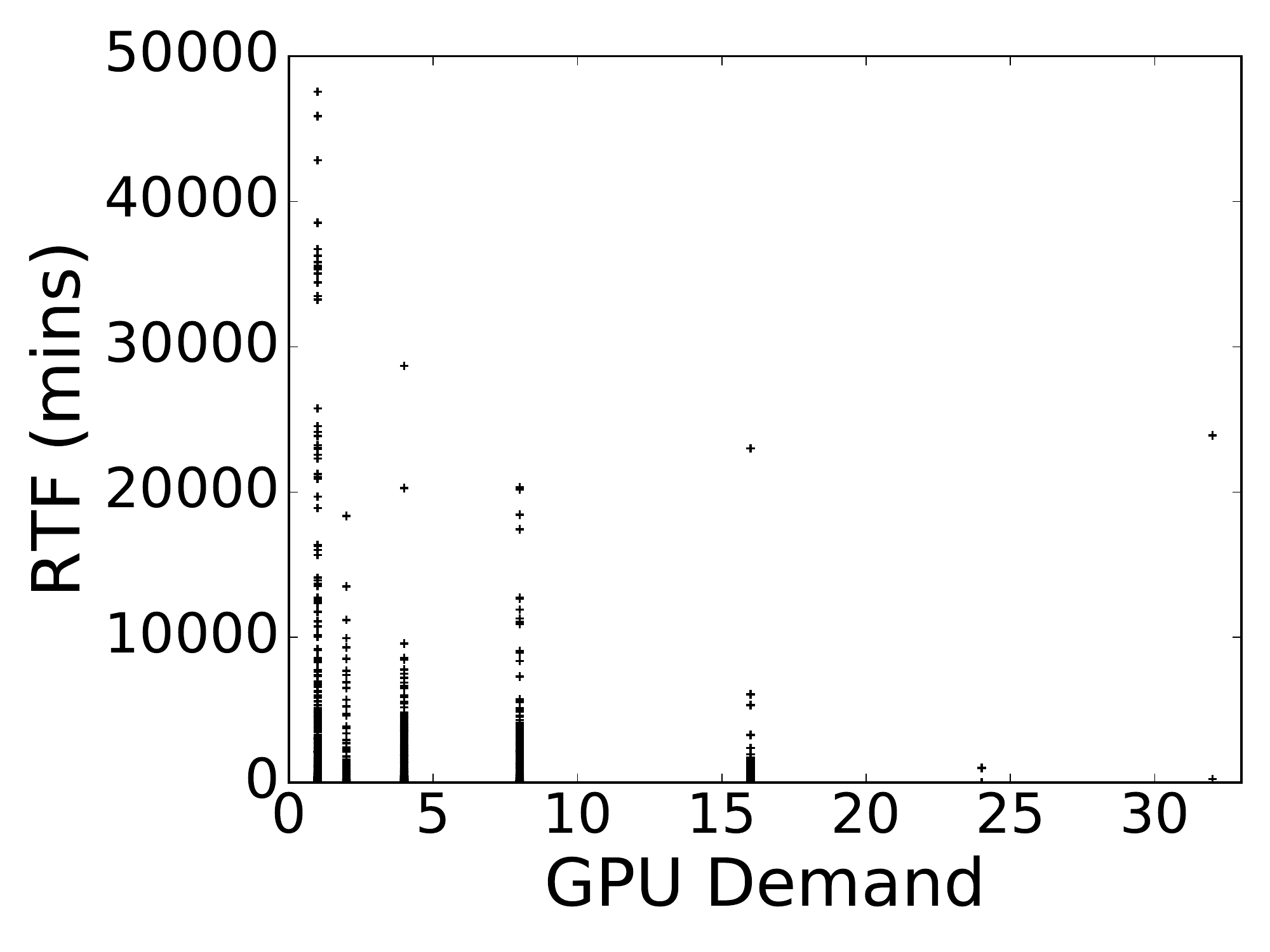}}
\subfigure[Semantic error]{\includegraphics[width=1.7in]{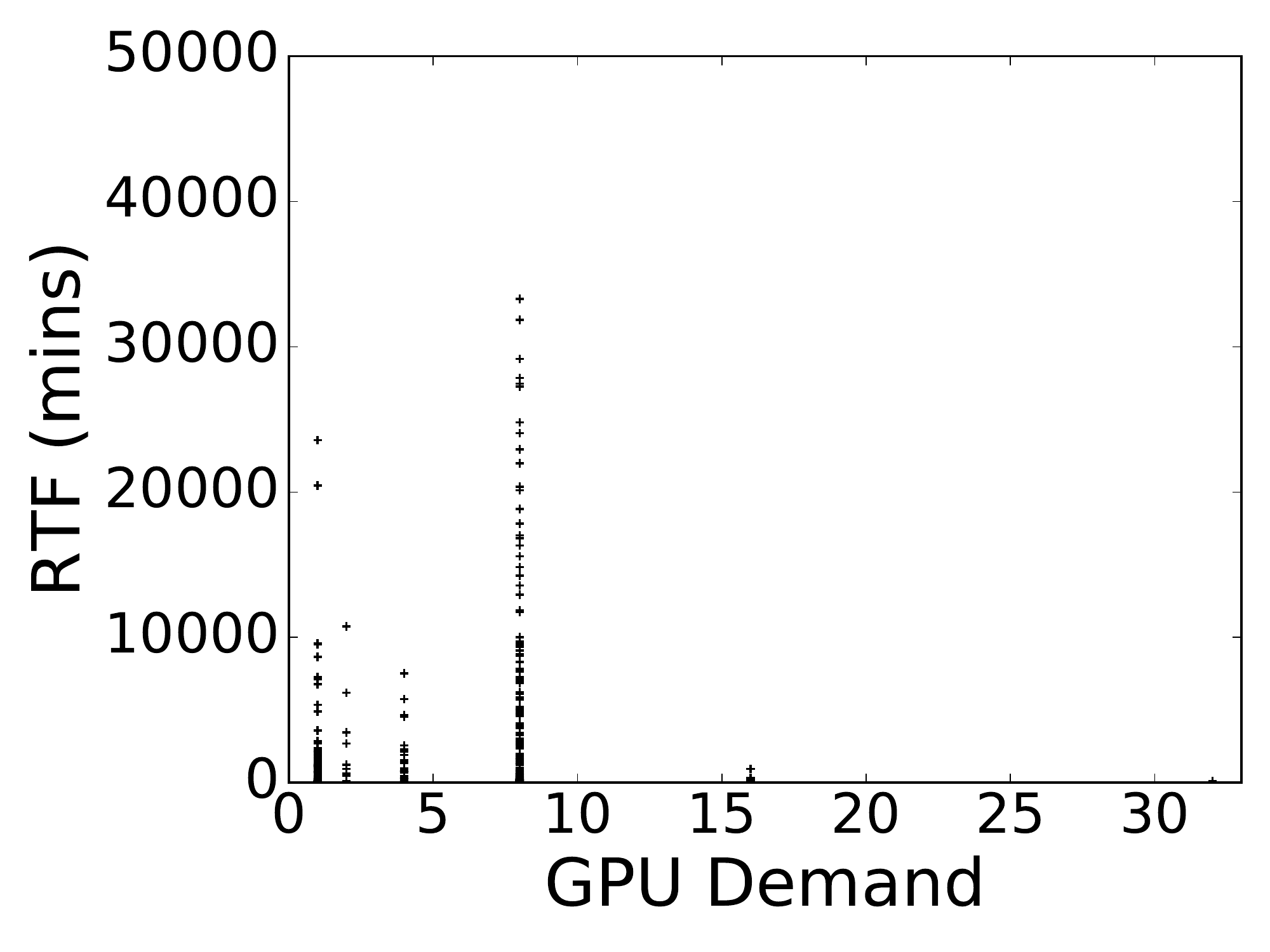}}\\
\subfigure[Model ckpt error]{\includegraphics[width=1.7in]{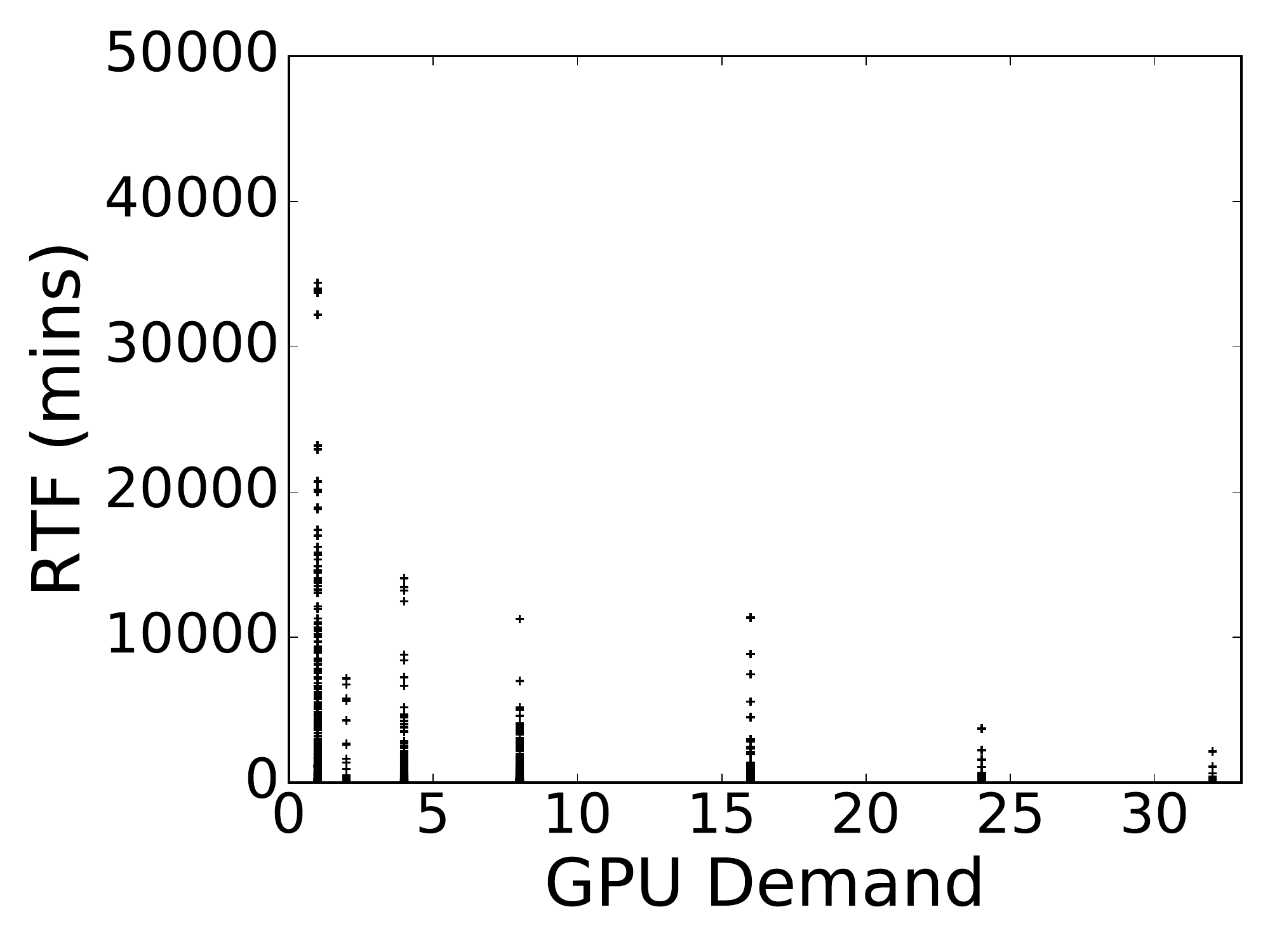}}
\subfigure[MPI runtime failure]{\includegraphics[width=1.7in]{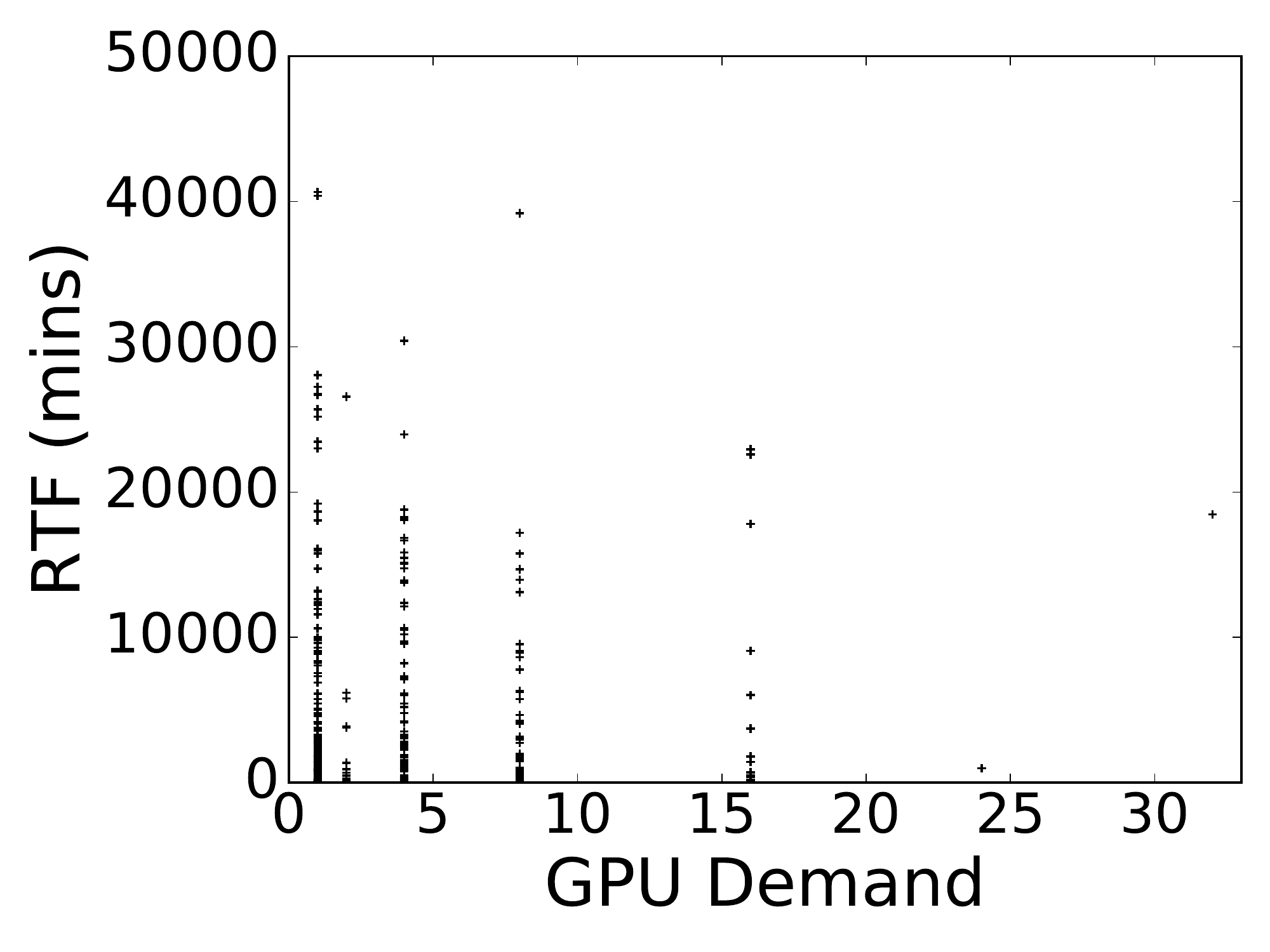}}
\end{adjustwidth}
\vspace{-0.1in}
\caption{\label{fig:demand_rtf_corr} Correlating RTF with GPU demand (i.e., number of GPUs) for four most RTF-dominant failure types.}
\vspace{-0.2in}
\end{figure}

\Paragraph{Large jobs with programming semantic errors tend to fail a while
after execution.} Column 5 of Table~\ref{table:failures} presents GPU demand
across failure reasons. To simplify the analysis, we select four
most-dominant failure reasons each contributing around 10\% or more in total
RTF. When we correlate RTF with GPU demand in
Figure~\ref{fig:demand_rtf_corr}, among the four failure types,
\emph{semantic error} exhibits a markedly distinct trend, with jobs that have
higher GPU demand
having relatively large RTFs, as compared to jobs having lower GPU demand.
This results in disproportional impact on the
actual resources utilized by failed jobs. We show this in Column 6 of
Table~\ref{table:failures}.

Column 6 presents actual GPU times for failures while multiplying RTF by GPU
demand. As the results show, compared to the RTF only, the impact of
\emph{semantic error} increases up to 17.06\% from 9.22\% while the other
three types of failure are either decreased or unchanged. This corresponds to
the fact that semantic error is relatively frequent in larger-demand
larger-RTF jobs. Looking deeper, we observe that training program instances
sometimes send, receive, and access data in an inconsistent way during model
parameters synchronization. As a consequence, \emph{semantic error} ranks the second
in terms of GPU resources used among failures in our workload.

\section{Design Implications for Future Schedulers}
\label{sec:Implications}

Based on our experiences and data-driven analysis so far, in this section we
discuss guidelines pertaining to the design of next-generation schedulers for
DNN training workloads.

\Paragraph{Prioritizing locality.} One of the main results from our analysis
of GPU scheduling was that lack of locality impacts both utilization and
 job running time. Our current scheduler adopts a classic approach where
it waits for a limited time to see if locality can be achieved and if not the
job is scheduled with the resources available at relaxed locality. The main
reason for this approach is to keep queuing time low as longer wait times
affect user experience.

However given that deep learning jobs run for many hours or even days,
incurring a 10\% or 20\% drop in efficiency could extend the job running time
by multiple hours. Thus in such scenarios, waiting for locality for a longer
time could be more beneficial. However this requires inferring long-running
jobs and appropriately setting user expectations. An alternate strategy could
be to \emph{migrate} a job to machines with better locality if resources
become available during the execution.



\Paragraph{Mitigating interference.} Another critical guideline for schedulers would be to
consider job placement policies to
mitigate inter-job interference. Instead of packing \textit{different} small
jobs on a single server, one option would be to place them on dedicated
servers, reducing sharing and thus interference among such jobs. Such an
option would increase fragmentation and will result in larger jobs having to
wait for longer if we have to prioritize for intra-job locality. Support for
job migration to \emph{defragment} the cluster~\cite{wencong1}, especially
applied to smaller jobs, will mitigate interference for small jobs, and will
improve intra-job locality for large jobs.



\Paragraph{Improving failure handling.} A large number of job failures we see
come from user errors in code or configuration. This is primarily because
programming languages in use are typically not strongly typed. We have found
that simple syntax checking could prevent many errors (e.g., missing
quotes or parenthesis) and some of the more sophisticated runtime failures
can be captured by running the first iteration of training.
We plan to set up a pool of cheaper VMs to pre-run jobs. Even running
multi-GPU jobs on a single GPU will catch such errors before they run on
larger shared clusters and thus prevent \textit{wasted} GPU cycles on them.
Training failures also happen due to erroneous data format (e.g.,
inconsistent columns in samples). We plan to investigate having a well
defined schema for datasets used in machine learning, and perform a schema
check while accessing data to reduce such errors.


Another useful extension for multi-tenant GPU clusters would be to develop a
system to predictively mitigate failures by proactively observing related
failures. The main goal of such a system would be to (i) classify error
messages in real time from logs that training jobs generate, and (ii)
adapting scheduling parameters per job (e.g., number of retries) as well as
across jobs (e.g., input data blacklisting) to reduce failure occurrences.
For example, the scheduler could stop retrying for failure categories like
incorrect inputs and continue retrying for network timeouts.

\section{Related Work}
\label{sec:RelatedWork}

\Paragraph{Failure analysis of data analytics jobs in shared clusters.} Prior
work has looked at designing large-scale data analytics platforms assuming
that failures are common~\cite{Barroso03,Dean13,Verma15}. They focus on
framework support for fault-tolerance and reliable job execution.
In this paper, we focus instead on understanding job failures in deep
learning specific platforms.

Kavulya~\textit{et al.} conducted a detailed characterization for job
failures in a production MapReduce cluster~\cite{Kavulya10}. Some of their
findings include: (1) Many jobs fail within a few minutes while the
worst-case job takes up to 4.3 days for its failure to be detected. These
failures occur due to data copy errors and are similar to HDFS-related
failures that we observe taking much longer to detect; (2) Many failures are
related to either exceptions due to array indexing errors or IO exceptions.
We again see some similarity to our work where coding errors lead to a number
of failure cases.



\Paragraph{Scheduler and runtime for efficient machine learning execution.}
SLAQ schedules concurrent machine learning training jobs based on quality
improvement for resource usage, allocating cluster resources for average
quality improvement~\cite{Zhang17}.
While this may improve
the quality across jobs, each individual job may take longer time to finish.
Optimus~\cite{Peng18} leverages the convergence curve to predict job
remaining time for dynamic resource scheduling and reduces average job
completion time. It adopts an online fitting model to derive a proper number
of servers and workers for MxNet~\cite{chen2015mxnet} jobs in parameter
server architecture. Tiresias~\cite{Juncheng19} reduces job completion times
when many training jobs undergo a \emph{trial-and-error} exploration where
job remaining time to complete training cannot be estimated from the
convergence curve. In this work, we found that a large job experiences highly
varying efficiency over placement spectrum (e.g.,
Table~\ref{tab:dist_degree}), and that future schedulers may need to consider
the trade-off between reducing queueing time and reducing job running time
more carefully over a wide range of locality choices.

\camera{We also note that an earlier technical report of our
work~\cite{Jeon18} was used to motivate new scheduling primitives in recent
work on scheduling like Gandiva~\cite{wencong1}. More importantly, our paper
presents a systematic study of a large-scale production cluster, covering the
whole lifecycle of deep learning jobs including queuing, execution, and
failure. We hope that our study of large clusters dedicated to deep learning workloads will
continue to motivate novel research in deep learning platforms and schedulers for these workloads.}

\Paragraph{GPU resource management for machine learning.} There are recent
efforts on GPU sharing for simpler machine learning tasks.
Baymax~\cite{Chen16} explores GPU sharing as a way to mitigate both queuing
delay and PCIe contention. Following that, Prophet~\cite{Chen17} proposes an
analytical model to predict performance of GPU workloads.
Gandiva~\cite{wencong1} proposes GPU time-sharing in shared GPU clusters
through checkpointing at low GPU memory usage of training job. Future work
includes integrating these prior work to improve cluster utilization and
capacity to run more jobs.

Many training
networks are memory bound, especially by capacity. Ryu~\textit{et al.}
analyzed memory allocation for ImageNet~\cite{ImageNet}, with recent VGG-16
model consuming up to 28~GB of memory~\cite{Rhu16}.
Therefore, vDNN~\cite{Rhu16}
proposes virtualized memory manager, and SuperNeurons~\cite{Wang18} adopts
fine-grained layer-wise memory control to schedule memory flexibly between
CPU and GPU.
Our work shares some similarity with prior findings (\textit{i.e.}, some
large networks do not fit in GPU memory) in real-world data.


\Paragraph{Approximate data processing.} Approximate data processing allows
trading off accuracy for earlier completion
times~\cite{Hellerstein97,Babcock03,Jermaine08,Condie10,Agarwal13,Zeng16}. In
databases, online aggregation has been studied in the context of SQL
queries~\cite{Hellerstein97,Condie10,Zeng16}. More recently, approximation
has been used in batch
processing~\cite{Agarwal13,Ananthanarayanan14,Venkataraman14}. Machine
learning training presents a fertile ground for exploring trading off
accuracy for early completion. In this paper, for the training workloads run
on our clusters, we quantify how trading off a very small amount of accuracy
(0.1\%) can result in significant savings in GPU execution time.


\section{Conclusion}
In this paper we analyzed a trace of deep learning jobs run on a large
multi-tenant cluster of GPUs and studied various factors that affect cluster
utilization. Our findings indicated the importance of locality for
distributed training jobs and also how interference from other colocated jobs
could lead to lower GPU utilization.
We also performed a detailed analysis of various failure causes and showed
how errors from various parts of the stack contribute to failures. Based on
our data analysis and experiences running a large-scale operation, we also
described guidelines that could help future research and development of
machine learning schedulers.

Finally, we have publicly released the scheduler trace containing information
about job arrivals, job size, placement and runtime to the community (details
available at~\cite{phillytrace}). As far as we know, this is the only trace
that includes rich information about deep learning training jobs run in
production. By making such a trace available, we hope to spur future research
in this area.


\section*{Acknowledgments}

We thank our shepherd, David Nellans, and the anonymous reviewers for their
valuable comments and suggestions. We also thank Lidong Zhou, Chris Basoglu,
Daniel Li, Marko Radmilac, Ashish Raniwala, Swapnil Palod and the rest of the
Microsoft Philly team for their unwavering help and support. This work was
supported in part by NRF-2018R1C1B5086586.

{\footnotesize \bibliographystyle{plain}
\bibliography{references}
}

\end{document}